\documentclass[prb,aps,superscriptaddress,showpacs,showkeys,reprint]{revtex4-1}
\usepackage{amsmath,amsfonts,amssymb}
\usepackage{bm}
\usepackage{natbib}
\usepackage{graphicx}
\usepackage{array}
\usepackage{verbatim}
\usepackage{multirow}
\usepackage{url}
\usepackage{bm}
\usepackage{hyperref}

\begin{document}

\title{Two-Component Coulomb Glass in Insulators with Local Attraction}

\author{Joe Mitchell}
\author{Anirban Gangopadhyay}
\affiliation{Center for Nanophysics and Advanced Materials, Department of Physics,University of Maryland, College Park, Maryland 20742-4111, USA}
\author{Victor Galitski}
\affiliation{Center for Nanophysics and Advanced Materials, Department of Physics,University of Maryland, College Park, Maryland 20742-4111, USA}
\affiliation{Joint Quantum Institute, Department of Physics, University of Maryland, College Park, Maryland 20742-4111, USA}
\affiliation{Kavli Institute for Theoretical Physics, University of California Santa Barbara, CA 93106-4030}

\author{Markus M\"uller}
\affiliation{Kavli Institute for Theoretical Physics, University of California Santa Barbara, CA 93106-4030}
\affiliation{The Abdus Salam International Centre for Theoretical Physics, P. O. Box 586, 34151 Trieste, Italy}

\begin{abstract}

 Motivated by evidence of local electron-electron attraction in experiments on disordered insulating films,  
we propose a new two-component Coulomb glass model that  combines strong disorder and long-range Coulomb repulsion with the additional possibility of local  pockets of a short-range inter-electron attraction. This model hosts a variety of interesting phenomena, in particular a crucial modification of the Coulomb gap previously believed to be universal. Tuning the short-range interaction to be repulsive, we find  non-monotonic humps in the density of states within the Coulomb gap. We further study variable-range hopping transport in such systems by extending the standard resistor network approach to include the motion of both single electrons and local pairs. In certain parameter regimes the competition between these two types of carriers results in a distinct peak in resistance as a function of the local attraction strength, which can be tuned by a magnetic field. 
\end{abstract}




\maketitle

\section{Introduction}

Disordered films with superconducting correlations host an amazing variety of interesting phenomena such as superconductor-insulator transitions tuned by disorder or an external magnetic field, with rather unusual transport properties~\cite{Hebard1990,Paalanen1992,Shahar1992,GantmakherInOxpeak,Sambandhamurthy2004,Shahar2005,Valles,Baturina2008,Sacepe2011,GantmakherDolgopolov}. These phenomena led to a lot of interesting theoretical work~\cite{Fisher, Finkelstein, Feigel'manKravtsov2010, Trivedi, GalitskiLarkin,Meir, MullerShklovskii, Nattermann},
which were to a great extent spurred by an experimental feature of many  strongly disordered films with superconducting correlations - a giant magnetoresistance peak - for which a full theoretical understanding is still lacking. Nevertheless, it was already understood in Ref. \onlinecite{Paalanen1992} that the presence of this peak, accompanied by Hall measurements, suggested the survival of some local pairing deep in the insulator, which is only gradually destroyed by an increasing magnetic field. Very recently, the presence of localized pairs has been verified by more direct STM spectroscopy \cite{Sacepe2011} showing the absence of coherence peaks in the tunneling density of states despite the presence of a superconducting gap -- a fact predicted theoretically earlier \cite{Trivedi}. The experimental evidence at hand support a distinct transition from a Bose insulating phase to a Fermi insulator and clearly require a detailed study of the strongly insulating regime which incorporates survival of the localized pairs.

A prominent material exhibiting the above phenomenology is InO$_x$, a commercially important and extensively studied semiconductor. 
Despite the uncertainties about its complex band structure, it is widely believed that the carriers in InO$_x$ originate from oxygen vacancies, likely partially compensated by the triply-negatively-charged indium vacancies. The recent ab initio study of Ref.~\onlinecite{Reunchan2011} calculated the formation energy of oxygen vacancies with different charge and found that a doubly-charged vacancy has the lowest formation energy in a crystalline environment (in zero field). The energetically next best state is an empty site, while a single occupied site corresponds to highly excited state. It is quite plausible that this tendency for local `pair' formation underlies the superconductivity in this system, similarly to compounds like PbTe~\cite{FisherStanford}, where local negative $U$ interactions have been proposed to lead to a non-standard type of superconductivity of preformed hard core bosons~\cite{Dzero2005}.

\vspace{0.00 in}
\begin{figure}[h]
\centering
 \includegraphics[angle=0, width = 0.53\textwidth]{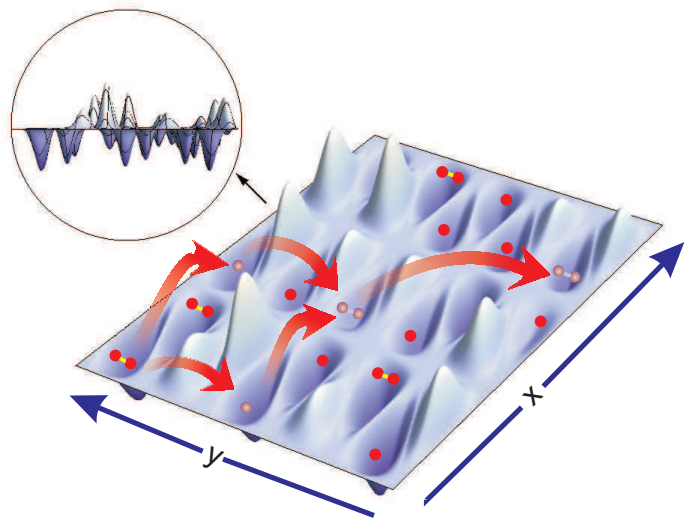} \\
 \vspace{-0.1 in}
 \caption{Illustration of the two component model: The energy landscape is due to the combination of on-site disorder and Coulomb interactions. The arrows indicate typical hopping processes relevant for the complex low $T$ transport in the two component Coulomb glass.  
 }
 \label{Illustration}
\end{figure}

A serious difficulty for the theory of such insulators stems from the need to treat strong disorder and electron pairing effects on equal footing.  In addition, 
recent experiments indicate that long range unscreened Coulomb interactions, often neglected in theoretical approaches, do in fact play an important role in several materials. In particular, the temperature dependence of the resistance in strong disorder~\cite{Shahar1992}, or on the high-field side of the mangetoresistance peak in TiN~\cite{Baturina2008} and InO$_x$~\footnote{B. Sac\'{e}p\'{e}, private communication.}, is often well described by the Efros-Shklovskii  law 
suggesting variable-range-hopping (VRH) in the presence of a Coulomb gap~\citep{ES1975}.  Although local pairing attractions, often captured through a negative-$U$ Hubbard model, have been studied quite extensively in previous research focussing on the superconductor-insulator transition ~\cite{Feigel'man2010, Dubi2006, Trivedi}, such studies have predominantly neglected long range Coulomb interactions. In contrast, we do include the latter and focus on more insulating regimes where Coulomb interactions play a crucial role and compete in a non-trivial way with the local negative $U$ attraction. Our study also has implications on Coulomb glasses in granular materials, where multiple occupation of sites is allowed. These aspects have recently been analyzed in closely related works~\cite{Chen2012,Chen2012a}. 
 



The above experimental motivations lead us to introduce a lattice model that  captures the various possible ingredients  present in the actual materials: 
strong disorder, local attraction of electrons (favored double occupancy of sites), and long-range Coulomb interactions, together with quantum transport captured by intersite nearest-neighbor hopping. The full Hamiltonian for such a system can be written in a general form, with the experimental tuning parameters (disorder strength $W$ and magnetic field $\bm{B}$) explicitly written,  as follows:

\begin{widetext}
\begin{equation}
\label{FullHamiltonian}
 \hat{H} = \sum_{i} \phi_i  \hat{n}_i + \frac{1}{2}\sum_{i,j} \frac{e^2}{r_{ij}} ( \hat{n}_i - \nu) (\hat{n}_j  - \nu) + \sum_{i} \frac{1}{2} U_i(\bm{B}) \hat{n}_i (\hat{n}_i -1 ) +  \sum_{\left\langle ij \right\rangle} \left( t^{(1)}_{ij} e^{\theta_{ij}(\bm{B})} \hat{c}_i^\dagger \hat{c}_j + 
 t^{(2)}_{ij} e^{2\theta_{ij}(\bm{B})} \hat{c}_i^\dagger \hat{c}_i^\dagger  \hat{c}_j \hat{c}_j + \text{h.c.}\right)
\end{equation}
\end{widetext}

In the above, $\phi_i=O(W)$ represents the random on-site potential due to the disorder, e.g. in the form of randomly positioned dopants. $ \frac{e^2}{r_{ij}}$ is the unscreened Coulomb repulsion between the localized carriers and  $U_i(\bm{B})$ is the local pairing interaction renormalized by the Coulomb repulsion between charges localized on the same site (within one lattice spacing). We assume that it is tunable, e.g., by the magnetic field. 
The last two terms represent quantum hopping of the single electrons as well as of pairs of electrons. 
While a pair hopping term is obviously generated as a second order process in single electron hopping, the relation between the single electron's tunneling amplitude $t^{(1)}_{ij}$ and that of the transfer of a pair may not be simple in the real materials, since it may involve details of the local electronic structure, which is responsible for the negative $U$ interaction. We therefore allow for a independent pair hopping amplitude $t^{(2)}_{ij}$. When we will discuss transport, we will take the two hopping amplitudes as independent phenomenological parameters, which translate into two independent localization lengths for localized single electron and pair excitations. The magnetic field enters the hopping terms via the phase factors $\theta_{ij}(\bm{B})$.

To solve this full quantum Hamiltonian would be an extremely ambitious goal. We will instead isolate individual aspects of this complex problem. In this work, we make two simplifying assumptions --- firstly, we focus on the regime in the phase diagram of these films where the  electron pairs are indeed formed locally, but are far from condensation. In technical terms, we treat the hopping terms in the Hamiltonian under the approximation $t^{(1,2)}_{ij} \ll {\rm max}(W, \frac{e^2}{a} )$ ($a$ being the lattice constant) and thus restrict ourselves to a classical model where transport is primarily through thermally-induced variable-range hopping, among exponentially localized states. This is closely analogous to the standard analysis of doped semiconductors \cite{ES1984}. When discussing variable-range hopping transport the hopping terms are taken into account via the (average) localization lengths, $\xi_1$ and $\xi_2$, of the single electrons and pairs, respectively, which are a result of the $B$-dependent hoppings $t^{(1,2)}$. Obviously, this approach prevents us from capturing superconductivity within the model. Nevertheless, many interesting physical phenomena observed in experiments, such as the giant magnetoresistance peak, often occur rather deep in the insulating phase ~\cite{GantmakherDolgopolov}, where such a strongly localized approach is meaningful.
 
Secondly, we assume that the entire effect of magnetic field is to tune the local pairing interaction.  It is reasonable to assume a monotonic decrease of the pairing strength $U$ with increasing  magnetic field.  However, in the present work we do not include the effects of the $B$-dependence in the hopping (orbital effects), but we focus entirely on the effect of changing the pairing interaction on various physical observables, such as the density of states and longitudinal resistance.  In an accompanying work \cite{Gangopadhyay}, we study instead the magnetic field dependence introduced by the phases in the hopping  terms in Eqn. \ref{FullHamiltonian} through explicit evaluation of the $\bm{B}$-dependence of the localization lengths $\xi_{1,2}$. In reality both effects are present simultaneously. We find that they both contribute to a non-monotonic magnetoresistance.
 
   

\section{Model}

We now focus entirely on this two-component Coulomb glass model, which will be shown to 
feature a significantly richer variety of phenomena than the canonical Efros-Shklovskii model. The latter considers a lattice of sites, $i$, with random on-site energies for electrons, $\phi_i$, populated with a filling factor, $\nu$. Each site $i$ can host only $n_i\in \{0,1\}$ electrons. The (classical) electrons repel each other with unscreened Coulomb interaction $e^2/r$, and the disorder is assumed to be distributed over a typical range $W$, e.g., uniformly in $\phi_i \in \left[-W,\, W\right]$.
An important hallmark of such systems is the soft Coulomb gap in the single particle density of states (DOS), $\rho(E)$, close to the Fermi level. For many materials with compensated doping, including InO$_x$, 
 the disorder is strong, i.e. $W \gg e^2/a$, where $a$ is the typical distance between neighboring electrons. In that case, the Coulomb gap is theoretically predicted~\cite{EfrosPikus,MullerPankov07} and empirically found~\cite{Mobius2009} to be essentially universal at low energies: 
$\rho(E)$ exhibits linear variation, $\rho(E)=\frac{\alpha}{e^4}|E| $. The co-efficient $\alpha$ is basically independent of the type of lattice, the filling fraction, and the details of the disorder~\cite{ES1987,Baranovskii1979}. We find a value $\alpha\approx 0.35\pm 0.01$ consistent with previous numerical studies~\cite{Mobius2009,Palassini11}, but substantially smaller than Efros' analytical estimate $2/\pi$ ~\cite{ES1984}.
The standard Coulomb gap shows up in transport as a stretched exponential resistance of the form 
\begin{equation}
\label{ES}
R(T)\sim R_0 \exp \left(\frac{T_0}{T}\right)^{\frac{1}{2}}.
\end{equation}
The co-efficient in the exponent, 
\begin{equation}
T_0 = C \frac{e^2}{\sqrt{\alpha}\xi_1}, 
\end{equation}
involves just one additional parameter: the average localization length, $\xi_1$, of single particle wavefunctions, apart from a numerical constant, whose value $4\lesssim C\lesssim 5$ can be extracted from a percolation analysis of random resistor networks~\cite{ES1984}, as well as from Monte Carlo simulations ~\cite{Galperin2012}. 
Now we extend the Efros-Shklovskii model by allowing double occupancy and electron pairing (cf.~ Fig.~\ref{Illustration} for an illustration) with the  classical Hamiltonian
\begin{equation}
 \label{Hamiltonian}
 H = \sum_i \phi_i n_i + \frac{1}{2}\sum_{j \neq i}\frac{e^2}{r_{ij}}(n_i-\nu)(n_j - \nu) +  \sum_i \frac{U_i}{2} n_i(n_i-1),
\end{equation}
where $n_i\in \{0,1,2\}$.  As mentioned above, the hopping will be reintroduced as a small perturbation to describe transport later.
The local attraction energies, $U_i$, for doubly occupied sites will be our control parameters driving the crossover from the electron-dominated regime ($U$ large and repulsive) to the pair-dominated regime ($U$ large and attractive). In between, we find a mixture of gapless single electron and pair states, which exhibits distinctly unique features that can be captured in experiments.  Note that the model (\ref{Hamiltonian}) is also of interest for  semiconductors in which doubly occupied sites (the upper Hubbard band) play a significant role~\cite{Ovadyahu-Kamimuraeffect, Vaknin96}. Many of the effects found here generalize in modified form to granular systems as well.

\section{Single site density of states}

\subsection{Definitions}
We start by analyzing the static properties of the two component electron glass. We consider the single site density of states (DOS) within typical metastable states. The latter are defined as classical occupancy configurations which are energetically stable with respect to moves of single electrons, pairs of electrons, as well as with respect to the formation of local pairs by combining two single electrons, or the reverse disintegration process. 
Let $S_n$ be the set of sites with occupancy $n\in \{0,1,2\}$  in such a local minimum configuration. 
We refer to the total energy to add (remove) a single electron on site $i$ as $E_i^{1+}$ ($E_i^{1-}$), and as $E_i^{2+}$ ($E_i^{2-}$) for pair excitations. We define the 
DOS for electron (pair) excitations, $\rho_{m=1(2)}$, as
\begin{equation}
\label{singleDOS}
 \rho_{m}(E) = \frac{1}{N} \sum_{i\in \Sigma^+_{m} } \delta(E - E_{i}^{m+}) + \frac 1 N\sum_{i\in \Sigma^-_{m} } \delta(E-E_{i}^{m-})
 \end{equation}
where $N$ is the number of lattice sites,  $\Sigma^+_1 = S_0 \cup S_1$, $\Sigma^-_1 = S_1 \cup S_2  $, $\Sigma^+_2 = S_0$, and $\Sigma^-_2 = S_2$. 

In the model without double occupancy ($U\to \infty$), imposing  stability  with respect to all possible single-electron moves,
\begin{equation}
 \label{ESStabilityConstraint}
  E_i^{1+} - E_j^{1-} - e_{ij} \geq 0,\quad e_{ij}\equiv \frac{e^2}{r_{ij}}, 
\end{equation}
is sufficient to induce the Coulomb gap in the DOS. Additional multi-particle constraints impose weaker conditions and have been shown to not significantly affect the low-energy profile of the DOS. In contrast, we show below that the presence of double occupancies results in additional constraints, which affect the Coulomb gap very significantly. In the following, we describe the evolution of the DOS as the attraction strength is tuned. 

Clearly, for strongly repulsive $U$, when all double occupancies are forbidden, the system reduces to the standard Efros-Shklovskii model where the canonical Coulomb gap with slope $\alpha \approx 0.35$ is found in the single particle density of states $\rho_1$.

\subsection{Spatially Uniform Interaction - Anomalous Coulomb gap}
\label{s:humps}

\begin{figure}
\centering
  \includegraphics[angle= 270, width=0.48\textwidth, totalheight=0.4\textheight]{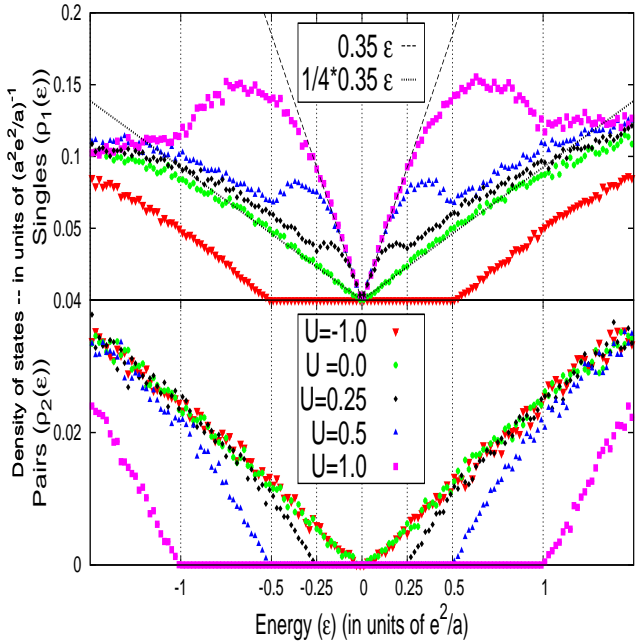}
 \vspace{-5pt}
 \caption{DOS for different uniform interaction $U$. $U=0$ is a critical point at which both $\rho_{1,2}$ have a linear pseudogap. The slope of the single particle DOS $\rho_1$ is suppressed to $\alpha/4e^4$. 
For net repulsion, $U>0$, $\rho_1$ has the canonical slope $\alpha/e^4$ at lowest energy, followed by a hump at the scale $U/2$, crossing over to the critical slope, while pairs are gapped up to $E=U$.
For $U<0$, single electrons have a hard gap $|U|/2$, while pairs are pseudogapped with slope $\alpha/16e^4$. Note: For these plots, the chemical potential was explicitly zeroed when averaging the DOS over the various initial occupancy-distributions}
 \label{DOSPlots}
\end{figure}

The case of a uniform pair interaction, 
$U_i = U \mbox{  } \forall i$,  can be understood essentially  analytically. Fig.~\ref{DOSPlots} illustrates the corresponding evolution of the DOS's with local interaction strength $U$, which were obtained from numerical simulations which we will describe below. In the attractive case, $U < 0$, all electrons remain paired in local minima, i.e. sites are either empty or doubly occupied. This is so because any singly occupied site would lower the energy be admitting a further electron brought in  from far away.
The pair-DOS, $\rho_2(E)$, is linear at low $E$, with the canonical slope $\frac{\alpha}{(2e)^4}$ corresponding to charges $2e$.  This results from the pair stability constraint analogous to Eq.~(\ref{ESStabilityConstraint}),
 \begin{equation}
    \label{StabilityConstraintPairs}
     E^{2+}_i - E^{2-}_j - 4e_{ij} \geq 0.   
 \end{equation}
This condition automatically ensures stability with respect to single-electron moves and pair formation/disintegration and thus constitutes the dominant condition determining the low-energy pair-DOS. 
These assertions are easy to check case by case, using that for $U<0$ single particle excitations are given by 
\begin{equation}
\label{sprelation}
E_i^{1\pm} = \frac{E_i^{2\pm}\pm |U|}{2}, \quad (U<0),
\end{equation}
on empty and occupied sites, respectively, being gapped up to energies $E_g=\frac{|U|}{2}$. As compared to a pair move, a single particle move does not only cost more in terms of onsite energy per particle, but also gives back less in terms of the polaronic interaction term $e_{ij}$. Likewise, one checks that if (\ref{StabilityConstraintPairs}) is satisfied, it is always unfavorable to let a pair disintegrate into two single electrons, partly because one loses the attraction energy $U$, and partly because one does not gain as much polaronic energy back.
The relation (\ref{sprelation}) implies that the single particle DOS is given by 
\begin{equation}
\rho_1(E) = 2\rho_2(2E - {\rm sgn}(E) | U|),  \quad (U<0).
\end{equation}
From this it follows that the single-DOS at energies beyond the gap and close to it goes as $\rho_1(E)= \frac{\alpha}{4e^4} (|E|-E_g)$. 


The point of no net interaction, $U = 0$, constitutes a critical point, where both $\rho_1(E)$ and $\rho_2(E)$ have soft excitations near $E=0$. However, most remarkably, the slope of $\rho_1$ is reduced by a factor of $4$ from its universal value $\alpha/e^4$ in the canonical model, as if it were the Coulomb gap of a system with effective charge $e^\ast=\sqrt{2}$e. This geometric mean of 2e and 1e arises because  
the gap imposed by the pair constraints (\ref{StabilityConstraintPairs}) is probed by 1e excitations. 
Indeed, for each pair of sites admitting a pair move, the constraint
  \begin{equation}
    \label{StabilityConstraintPairs2}
     E^{1+}_i - E^{1-}_j - 2e_{ij} \geq 0
 \end{equation}
 must hold, as one obtains by inserting (\ref{sprelation}) for $U=0$ into (\ref{StabilityConstraintPairs}).
 This is indeed a more stringent constraint than Eq.~(\ref{ESStabilityConstraint}).

On the repulsive side, $U>0$, pairs are gapped up to energy $E_g = U$. Mathematically, this follows simply from the fact that on empty sites, one has 
  \begin{equation}
  \label{E2lambdapos}
E^{2+}=2E^{1+}+U, \quad U>0,
\end{equation}
with $E^{1+}>0$, and an analogous relation for doubly occupied sites. Indeed, to accomodate a pair in a potential well, the well must be at least as deep as $-U$, which ensures that  the second electron is just loosely bound. The minimum energy required to remove the pair from such a well is $U$. Similarly, injecting a pair into an empty site costs at least the repulsion $U$ of the second electron.

On the other hand, for repulsive $U$, $\rho_1(E)$ remains ungapped. At low energies ($ |E| \ll \frac{U}{2}$), the universal single-electron Coulomb gap with slope $\frac{\alpha}{e^4}$ emerges: indeed, the vast majority of stability constraints involving sites at these energies are single-electron constraints.
At larger energies, $|E|\gg U$, one can ignore $U$ in the stability constraints, which then reduce again to Eqs.~(\ref{StabilityConstraintPairs},\ref{StabilityConstraintPairs2}) and thus lead to a slope of $\frac{\alpha}{4e^4}$ (for $E$ below the Coulomb gap, $E_{\rm Cb} \sim \frac{(e^2/a)^2}{W}$).   
This immediately leads to an interesting prediction: in the repulsive case, at intermediate energies, $U/2\leq |E| \leq U$, $\rho_1(E)$ is non-monotonic, as is indeed confirmed by the numerical data in Fig.~\ref{DOSPlots}. 

Let us now characterize the single-DOS in this regime in more detail. At small positive $E\ll U$, $\rho_1(E)$ receives essentially equal contributions from empty and singly occupied sites. Likewise, for negative energies, it receives equal contribution from doubly and singly occupied sites . If we denote the respective contributions as  $\rho_1^{(0)}(E)$, $\rho_1^{(1)}(E)$ and $\rho_1^{(2)}(E)$ (with superscripts denoting occupancies), we find empirically that 
 \begin{eqnarray}
 \label{rho1_poslam}
\rho_1^{(1)}(E)& \approx & \rho_1^{(0)}(E)\approx \frac{\alpha}{2}\frac{ E}{e^4}, \quad 0<E \ll U,\nonumber \\
\rho_1^{(1)}(E) & \approx & \rho_1^{(2)}(E)\approx \frac{\alpha}{2}\frac{ |E|}{e^4}, \quad 0<-E\ll U,
\end{eqnarray}
i.e., the ground state occupation is practically uncorrelated with the excitation energy.  

\begin{figure}
 \includegraphics[angle=270, width = 0.45 \textwidth]{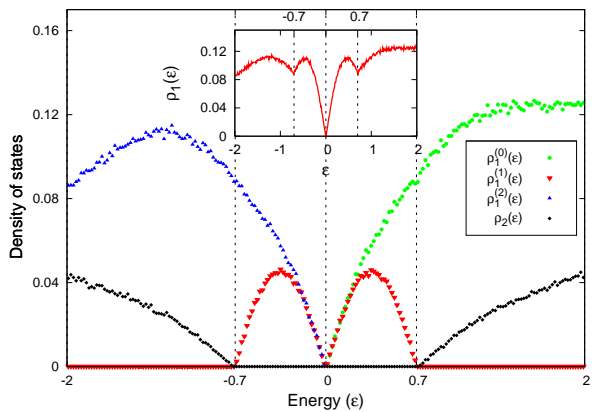}
 \vspace{10pt}
   \caption{Breakup of the single-particle DOS $\rho_1(\epsilon) = \rho_1^{(0)}(\epsilon) + \rho_1^{(1)}(\epsilon) + \rho_1^{(2)}(\epsilon)$ for repulsive $U = 0.7$, split according to the site occupancies, as described by Eqns. \ref{rho1_poslam}, \ref{rho2_poslam} and \ref{rho2_poslam2}. Note that the contribution to $\rho_1(\epsilon)$ from singly-occupied sites ends at $\epsilon = \pm U$, exactly where the pair-DOS $\rho_2(\epsilon)$ begins. Inset: The dip in $\rho_1(\epsilon)$ corresponds to $\rho_1^{(1)}(\epsilon)$ going to zero at $\epsilon = \pm U $, as emphasized by the dashed lines. }
   \label{DOSBreakup}
\end{figure}

From Eq.~(\ref{E2lambdapos}) it follows that the pair DOS $\rho_2(E)$ satisfies
\begin{equation}
\label{rho2_poslam}
\rho_2(E) = \frac{1}{2}\rho_1^{(0)}\left( \frac{E-{\rm sgn}(E)U}{2} \right).
\end{equation}
Thus, from (\ref{rho1_poslam}),  beyond the pair gap it starts off as
\begin{equation}
\label{rho2_poslam2}
\rho_2(E) \approx  \frac{\alpha}{8}\frac{|E|-U}{e^4},
\end{equation}
as can be seen in Fig.~\ref{DOSBreakup} 

Note that the contribution to $\rho_1(E)$ from singly occupied sites is restricted to the energy range $|E|\leq U$, since otherwise spontaneous particle rearrangements  would occur. Further, $\rho_1^{(1)}(E)$ satisfies the simple relation
  \begin{eqnarray}
  \label{extraconstraint}
\rho_1^{(1)}(E) = \rho_1^{(1)}(E-U), \quad 0<E< U,
\end{eqnarray}
which expresses the relationship $E^{1+} = E^{1-}+U>0$ between particle addition and removal for singly-occupied sites. Note that this implies in particular that 
\begin{equation}
\rho_1^{(1)}(|E|\to U)\approx \frac{\alpha (U-|E|)}{2e^4}
\end{equation}
tends to zero at $E=\pm U$, and has a maximum around $E=\pm U/2$.

 At the same time the contributions $\rho_1^{(0,2)}(E)$ to the single-DOS do not exhibit any sharp features at energies of order $U$, except that they smoothly roll-over from a slope $\frac{\alpha}{2e^4}$ at $|E|\ll U$ (cf. Eq.~(\ref{rho1_poslam})), to a slope that approaches $\frac{\alpha}{4e^4}$ for  $|E| > U$. As a result, the full single particle DOS, which is the sum of these two contributions, exhibits a local maximum around $E= \pm U/2$ and a local minimum around $E=\pm U$, essentially reflecting the properties of $\rho_1^{(1)}(E)$ imposed by the extra constraint (\ref{extraconstraint}). Very similar physics was found recently in granular systems ~\cite{Chen2012} where the occupancy of sites is nearly unlimited, in which case Eq.~(\ref{extraconstraint}) applies essentially to the whole 1-particle DOS, and imposes mirrored Coulomb gaps.

Despite the absence of quantum fluctuations, the described evolution of the DOS has a lot in common with quantum critical phenomena ~\cite{Sachdev2000}, where $U$ plays the role of the detuning parameter from criticality. The critical behavior is restored at energies $|E|\gg |U|$, with linear DOS-s and anomalous slope of $\rho_1(E)$. At low energies, the non-critical phase appears, where one type of carriers is gapped out, while the other type exhibits a universal Coulomb gap.
We also note that the features of the DOS and the underlying mechanisms found here have similarities with those in a recently proposed model of strongly and weakly interacting two-level systems~\cite{SchechterStamp2009}.

\subsection{Numerical simulations}
In order to analyze further details of the DOS, as well as the case of random local interaction $U_i$, we performed numerical studies.
To study metastable states, we start from a random configuration of occupancies, $n_i$ ($\in \left\lbrace0,1,2 \right\rbrace$), on a half-filled triangular lattice of size 200 $\times$ 200. We choose a triangular lattice with commensurate filling so as not to introduce extra strain in the system in the limit of weak disorder (note that if the filling is not commensurate there is still some strain from the lattice). 
However, we focus on strong disorder where the effect of the lattice type is expected to be small. 

Following a similar protocol as described in \cite{Baranovskii1979}, we allow re-distribution of occupancies through single particle moves, pair moves and pair dissociation/formation --- the last one within a restricted spatial range --- that lower the total energy of the system until the system stabilizes in a local minimum of the energy. In this context, it is important to recall that  the appearance of the Coulomb gap in the single particle DOS does not require stability with respect to multi-particle moves, and is not very sensitive to the latter. This is because the single-particle moves impose the strongest stability constraints~{\cite{Baranovskii1979}}. By a similar reasoning, the universal features in the DOS for uniform $U$ in the model considered here result from single particle and pair stability constraints. It is thus reasonable to expect that the class of moves considered above imposes the strongest stability conditions, determining the essential features of the single site DOS-s, $\rho_1(E)$ and $\rho_2(E)$, for single electron and pair excitations, respectively.
Further multiparticle processes  may relax the system to lower lying metastable states; however, such states are expected to have very similar single site density of states and transport properties.  The single site DOS was obtained by calculating the histogram of the energies  to add or subtract an electron or pair from each site, cf.~Eq.~(\ref{singleDOS}).
These DOS-s were averaged over many different disorder realizations, typically of the order of $100$ for the $200\times200$ sized systems.

We measure all distances in units of the lattice constant $a$, and in our finite-sized samples, the intersite distance $r_{ij}$, has been chosen as the minimum distance on a torus defined by periodic boundary conditions.
Energies are measured with reference to the chemical potential $\mu$ in units of the nearest neighbor Coulomb repulsion $\frac{e^2}{a}$. The chemical potential in this case is determined as the average of the smallest energy to add and remove an extra particle from a given metastable state. Pair energies are measured from the reference energy $2\mu$.

We choose the on-site disorder $\phi_i$ to be randomly distributed in the interval $[-W,W]$. It is well-known that a disorder of order unity or more is required for the DOS to tend to an essentially universal Coulomb gap, $\rho_1(E)=\frac{\alpha |E|}{e^4} $  at low energies.  In our model, since a site is allowed to have double occupancy, the strong disorder condition is met  when $W$ exceeds the typical nearest neighbor interaction of two doubly occupied sites. In order to find DOS features which approach a universal limit, we therefore chose to work with disorder $W=4$. At substantially weaker disorder the low energy DOS's were  indeed found to be non-universal.  

\subsection{Spatially Disordered Interaction}

\begin{figure}
 \includegraphics[angle=270, width = 0.45 \textwidth]{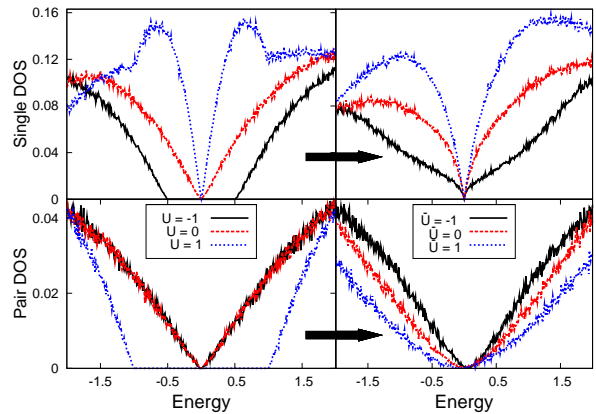}
 \vspace{10pt}
   \caption{Comparison of the DOS's for constant (left) and random (right) $U$, with strong scatter $\Delta U = 2$ (units of $e^2/a$). The sharp gaps and humps are smoothed out by disorder, but the overall trend of increase/decrease of the low energy DOS remain intact.}
   \label{Constant2RandomLambda}
\end{figure}

In a disordered system, it is more realistic that the pairing energies $U_i$ are non-uniform. 
Fig. \ref{Constant2RandomLambda} shows the result for the DOS-s for a model  with  $U_i$ distributed randomly in an energy range $\left[ \overline{U}-\Delta U, \overline{U} + \Delta U \right]$.
The sharp features of Fig.~\ref{DOSPlots} are smoothened. The gaps in $\rho_1(E)$ and $\rho_2(E)$, for repulsive and attractive $\overline U$ respectively, are smeared out. Low energy single particle states leak into the gap of $\rho_1(E)$ as soon as there are positive $U_i$-s. The density of such states grows with increasing $\overline U$, and eventually saturates to the standard linear pseudogap with slope $\frac{\alpha}{e^4}$. Closely analogous considerations apply to $\rho_2(E)$ upon decreasing $\overline U$. The detailed behavior in the intermediate regime $|\overline U| \leq \Delta U $ is complicated and presumably non-universal. 
 

The non-monotonic humps in $\rho_1$ discussed in section~\ref{s:humps} survive only if $\Delta U$ is sufficiently small as compared to $\overline U$. Thus they are probably best sought after in crystalline samples, where the local environment of different impurities are similar, giving rise to a narrow scatter, $\Delta U$.


\section{Transport and Resistance}

\subsection{Choice of parameters}

In an insulator, a reduced density of states is usually reflected in     an exponentially increased resistance. It is thus interesting to ask what happens in the "mixed regime" of our model, where both pair and single electron excitations are ungapped. 
If transport was dominated by one type of excitation only,
one would expect an increase of resistance upon approaching the mixed regime from either side, since the DOS of the dominant carrier type diminishes. However,
transport is more complicated in this two-component Coulomb glass. Electron and pair hops do not take place independently in the sample, but combine to form a network of interconnected pair and single electron moves, as illustrated in Fig.~\ref{Illustration}. Transport is a complex functional of the {\em combined} density of states.
In order to study this insulating regime, we have generalized the construction of an effective network of Miller-Abrahams resistors~\cite{MillerAbrahams1960} which include  both pair and single particle processes.  In order to elucidate the interplay between pair and single particle transport, we have neglected spin blocking effects in the random resistor network~\cite{Ovadyahu-Kamimuraeffect} (as may be justified in strong spin orbit coupled materials). Spin dependent effects may be considered elsewhere. The elementary hopping resistances were evaluated in a mean field fashion for selected metastable states~\cite{ES1984, Amir2009}. 

Details of the resistor network, the required steps and approximations, are described in the appendix. The only important point to note is the exponential dependence of the effective resistances of the network on temperature and localization length.  This restricts the accessed range of energies and typical hopping distances of the electrons participating in the network and using a percolation argument~\cite{Ambegaokar1971}, one can determine the functional dependence of the resistance on temperature.

In presence of the Coulomb gap in the density of states, and if one type of carriers dominates the low $T$ transport, the functional dependence is of Efros-Shklovskii type, cf. Eq.~(\ref{ES}), with 
\begin{equation}
\label{Efros-ShklovskiiTemperature}
T_0^{(i)} =  C \frac{Q^2_ie^2}{\sqrt{\alpha}\xi_i}, 
\end{equation}
where $Q_{i=1,2}$ is the charge of the carriers in units of $e$, $\xi_i$ their average localization length and $C\approx 4-5$.  
These localization lengths may in principle be evaluated from an analysis of the elementary localized excitations above the ground state, whose spatial extent is governed by the hopping terms in Eq. \ref{FullHamiltonian}. The magnetic field enters this localization length via phases in the hopping and resulting interference effects, as discussed e.g. in Ref.~\cite{Zhao1991,Mueller}. However, for the purpose of analyzing the effect of varying pairing strength, we assume the localization lengths to be constant. This may describe very well an experimental situation in which the local interaction $U$ is tuned (by chemical modifications or gating), without affecting the localization lengths. In contrast, in the case where $U$ is tuned by a magnetic field, the effects described below will necessarily be superposed over quantum interference effects, which affect $\xi_{1,2}$ rather sensitively and may well dominate the effects which we address below.
  
As we shall explain below, under certain circumstances we obtain a  nonmonotonicity in resistance as a function of $U$. The latter is most prominent when we have $\frac{\xi_2}{\xi_1}=4$, in which case the Efros-Shklovskii temperatures $T_0^{(1,2)}$ are the same for both single-electron and pair-transport resulting in an interesting competition when $U$ is tuned across zero. 
In reality the ratio $\xi_2/\xi_1$ varies greatly across the phase diagram of disordered films with superconducting correlations. Indeed
the localization length of preformed pairs must diverge at the transition to a superconductor, while $\xi_1$ remains non-critical~\cite{Anderson59, MaLee85}. On the other hand, far in the insulating regime, $\xi_2$ is expected to  become shorter than $\xi_1$, because pair tunneling is suppressed.
Therefore a regime where $\xi_2 > \xi_1$ should certainly exist, and below we consider the particularly interesting case $\xi_2/\xi_1 = 4$. We should keep in mind however, that such a large ratio presumably implies relatively strong quantum fluctuations, due to rather important hopping terms. With this caveat in mind, the essentially classical description of the two-component Coulomb glass presented here should be taken as a phenomenological approach to capture Coulomb frustration effects on a system with variable range hopping transport of competing carriers. 


\subsection{Results}

In the case of uniform $U$, there is no genuine mixed transport regime at low $T$, since one of the two carrier types is always gapped.
Here we find the resistance, $R(U;T)$, to be a flat, i.e., constant function of $U$ within  error bars. Despite a suppression of $\rho_1(E)$ (see Fig.~\ref{DOSPlots}), the resistance does not increase significantly. As $U$ is decreased, the pairs ``fill in" the resulting gap left in the transport channels, keeping the resistance essentially constant. 

\begin{figure}[h]
\centering
 \includegraphics[angle=270, width = 0.48\textwidth]{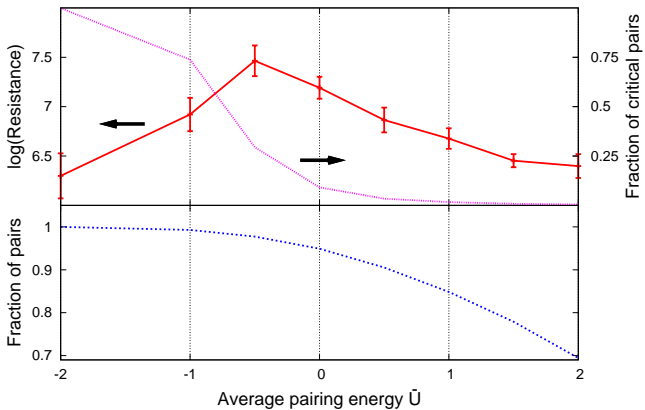} \\
 \vspace{-8pt}
 \caption{Top panel: Peak in the resistance upon tuning $\overline{U}$ at $T=0.04$ vs. the average interaction $\overline{U}$ (with $\xi_2=4 \xi_1$); and the fraction of pair hops in the percolating cluster (with resistances within $30\%$ of the percolating resistance). The resistance peaks when roughly half of the critical resistors are pair and single moves, resp. 
Lower panel: fraction of paired electrons in a typical metastable state. This fraction smoothly decreases across the "mixed regime", since the bulk of such pairs is inactive in transport.
 }
 \label{ResistancePeak}
\end{figure}

However, the situation is more interesting with random $U_i$, where a genuine mixed two-component carrier regime exists. If pairs and single-electron excitations have strongly disparate localization lengths, $R(\overline{U})$ is still monotonic under tuning of $\overline U$, essentially reflecting the evolution of the DOS  of the less localized carrier type. However, the two carriers do compete significantly in an intermediate regime. Indeed, we find an interesting non-monotonicity in the resistance; see the top panel in Fig. \ref{ResistancePeak}.  At low temperatures
we find a relatively significant peak in the resistance, centered around a small $\overline U < 0$. This feature is even slightly enhanced by increasing the randomness $\Delta U$.
Transport in the peak region is partially by pairs, which break up and propagate as single electrons, and then recombine again. The maximum of resistance occurs when roughly an
equal number of single and pair hops form the critical links of the percolation network, see Fig.~\ref{ResistancePeak}.

A plausible qualitative explanation for the numerically observed peak is the following: it is difficult to connect regions in which pair or single electron transport is favored, as opposed to the regimes $|\overline{U}|\gg 1$, where transport is dominated by one type of carrier only. In other words, the mixed regime suffers from ``contact resistances" between pair- and single-dominated parts of the resistor network in the following sense. A piece of transport path of pairs, must be connected to {\em two} good strands of single particle transport, and likewise a single particle transport path must find another one to continue as a pair path. Both links require some matching which tends to increase the overall resistance. This phenomenological explanation of our numerical observations bears some resemblance with the idea that superconducting islands may act as weak links in a single-electron-dominated transport regime on the insulating side of the SIT, as proposed in \cite{Dubi2006}.


\section{Conclusion}

The model discussed in this paper, should be realized in disordered materials with a strong tendency for local attractions (negative Hubbard $U$). It also predicts interesting effects for cases where local interactions are moderately repulsive, such that multiple occupancy of sites is still possible. The occurrence of negative $U$ interactions is likely to be concomitant with a bosonic type of superconductor-to-insulator transition upon further reduction of the disorder. In such samples, $\overline U$ may also be tuned by an external magnetic field which has a depairing effect on the electrons. However, since a magnetic field also sensitively affects localization lengths, it would be desirable to use other, non-magnetic means to influence the local interactions, too (such as pressure, chemical doping etc). 
If the disorder in the local $U$-s is large we find a regime around $\overline U\approx 0$, where both pairs and single electrons contribute to the activated transport, and a non-monotonic resistance as a function of $U$ results.

For the strongly localized, classical limit of the two-component Coulomb glasses we found several interesting effects on the low energy density of states. In particular, we find that the tendency for local attraction leads to a suppression of the density {\em beyond} the standard Coulomb gap. At the point where local attraction and repulsion balance to produce vanishing net interaction $U=0$ we find that the 2d Coulomb gap is reduced by a factor of $4$ from its canonical value. More generally, if multiple charging of the same site with $M$ charges (without paying additional local charging energy) were allowed, one would find a suppression by a factor of $M^2$.  

For the case of moderately repulsive interactions $U>0$, if the randomness in the interaction energy $\Delta U$ is small compared to the average interaction $\overline{U}$, our model predicts the existence of non-monotonic humps in the single particle density of states. If pair transport is suppressed due to strong localization, the non-monotonicity of the single-electron DOS in the repulsive case $U>0$ should show up as a kink in the resistance $R(T)$ around a temperature $T_*\approx (U/2)^2/(Ce^2/\xi)$, where it crosses over from an Efros-Shklovskii law with a higher value of $T_0$ to a less steep $R(T)$ and a twice smaller $T_0$ at lower $T$. The humps in $\rho_1$ should also leave traces in AC measurements~\cite{Armitage}, or more direct measurements of the DOS such as photoemission or tunneling from a broad junction~\cite{MasseyLee}. These DOS features may also be relevant for the more involved experiments of memory effects in deep insulators~\cite{electronglass,Lebanon2005}, where doubly occupied sites with repulsive interactions are known to be present~\cite{Vaknin96}. 

A measurement of the pair-DOS, especially on the attractive side $U<0$ could be attempted through measurement of the tunnelling conductance from a (wide) superconducting probe, similar to the experiments performed by Dynes et al \cite{Naaman2001}.

In this work we have taken the localization lengths to be independent of the tuning of the local interaction strength. If the latter is tuned by magnetic field, a full description needs to take such quantum effects into account, however.  
In fact, it has been argued~\cite{Zhao1991, Mueller}  that the  field dependence of localization lengths of pairs and electrons are opposite, which is probably an important ingredient for a strong magnetoresistance peak. Here we show that, on top of that effect, the complex energetics and transport phenomena in the two component Coulomb glass can even enhance such a peak. A discussion of the combination of both magnetic field effects (tuning $U$ {\em and} the localization) is left for future studies.

\section*{Acknowledgments}

This work was supported by the Department of Energy (AG and VG) and NSF-CAREER award (JM and VG). 
and by the National Science Foundation under Grant No. NSF PHY05-51164. 
The authors thank B. Shklovskii and B. Sac\'{e}p\'{e} for helpful discussions.

\section*{Appendix A : Resistor network mapping}

The problem of finding the conductivity of a sample, where transport takes place via variable range hopping of electrons from singly occupied sites to empty sites, can be mapped to an equivalent random resistor network problem (see Ref. \onlinecite{MillerAbrahams1960} for a derivation). The sites in the hopping problem can be mapped to the vertices of this network, and the inter-site transition rates to the resistances linking these vertices.  Finding the sample resistance simply reduces to calculating the effective resistance of the resistor network through a percolation approach ~\cite{Ambegaokar1971} .  

It is possible to formulate a similar resistor network mapping for variable range hopping on a lattice where double occupancy is also allowed. Early efforts in this direction were made by Kamimura et. al. \cite{Ovadyahu-Kamimuraeffect}. There, the  inter-site transition rate was taken as the sum of transition rates of the four types of possible single-particle processes characterized by the occupancies of the initial and final sites before the hop.

  In our two-component model, we additionally  introduce the pair-hopping channel between two sites alongside the single-particle transport channels considered in Ref. \onlinecite{Ovadyahu-Kamimuraeffect}: in what follows, we refer to these single particle and pair hops as first order processes. However, with this introduction, two-particle processes involving more than two sites may also become significant as they can potentially provide lower-resistance alternatives to two-site pair hops between given initial and final occupancy configurations.  For brevity, we refer to these two-particle processes involving more than two-sites as second order processes with the idea that two such processes in succession (a pair breaking followed by re-joining of the lone electrons) can constitute a first order process (a pair hop). The  book-keeping of the different allowed hopping processes based on occupancies prescribed in Ref. \onlinecite{Ovadyahu-Kamimuraeffect} becomes increasingly more cumbersome, as one takes into account these second-order processes.

Hence we formulate a more general prescription for the resistor-network mapping which can be naturally extended to include  higher-order processes.  We first describe this reformulation for the case where transport takes place through first-order processes only. In this case, each vertex of the equivalent resistor network corresponds to  a node $(i,n_i)$ defined as a site $i$ together with its occupancy $n_i\geq 1$ . In a system with double occupancies allowed each site gives rise to two nodes, while upon eliminating double occupancies the only nodes are $(i,n_i=1)$, which reduce to the standard Miller-Abrahams network. 
 The various first order processes between the two given sites $i$ and $j$ correspond to resistances between different nodes of the network: see Table \ref{Resistors} for a full description of the four possibilites. In networks without double occupation, there is only one resistance, $R(i,1;j,1)$ associated to a given pair of sites. 
 
 In defining the resistances, it must be noted that, while the occupancies of the two sites under consideration  assume all possible nonzero values, the occupancies of all other sites are frozen at a pseudoground state.

  The expression for the transition rate given below in Eq. (\ref{transition}) for a single-particle hopping process of one type tallies exactly with that of Ref. \onlinecite{Ovadyahu-Kamimuraeffect}. Since it is expected that, between two given
sites, one type of single-hopping process is energetically favoured, and thus has least resistance compared to the other three, the two resistor network mappings might be naively expected to produce identical results if only single-particle hops are considered. However, a conducting path through the sample constructed from these least resistance hops might involve an occupancy mismatch between resistors sharing the same site: the  alternative paradigm for the resistor-network construction in our approach, which includes occupancy in the definition of vertices of the network, excludes this possiblity.
  
  In the second section, we will include second-order processes by a simple extension of the definition of nodes from $(i,n_i)$ to $(i,n_i,j,n_j)$. Herein lies the utility of our approach : higher-order multi-particle processes can be easily incorporated simply by extending the dimensions of the node-network. The resistor network constructed from these generalized nodes contain the first-order processes as a subnetwork as will be described in detail later. This extension, however, makes calculation of the effective resistance very expensive. Therefore, in the last section, we describe possible approximations so that we can investigate the effect of pair breaking/formation (crucial two-particle moves in the regime where both single and pair DOS's are ungapped) despite staying within a simpler resistor-network with nodes of the form $(i,n_i)$.

\subsection*{1. Resistor-network construction for first order processes}

Let us now describe the resistor network mapping in detail. We start by reaching the pseudo-ground state, used earlier to extract the density of states.

In zero field, the time-averaged rate of transfer of electrons through single-hops from node $(i,n_i)$ to node $(j,n_j)$ (note: $n_i$ is the occupancy of site $i$ \emph{before} the hop, while $n_j$ is the occupancy of site $j$ \emph{after} the hop) is given by 
   \begin{equation}
    \label{transition}
    \Gamma^0(i,n_i;j,n_j) =  e^{-\frac{2r_{ij}}{\xi}}   \frac{e^{-\beta E(n_i,n_j-1)} }{Z}   P(i,n_i;j,n_j) 
   \end{equation}

    Here $ r_{ij} = \left| r_j - r_i \right|$ is the distance between the sites and $\xi$ is the localization length of the electronic wavefunctions for single-hops and that of the pairs for pair hops. $E(n_i,n_j-1)$ is the total energy of the  system when sites $i$ and $j$ have occupation numbers $n_i$ and $n_j-1$, respectively (for all sites other than $i$ and $j$, we use the pseudo-ground state occupancies).  Thus, the term $\mathrm{e}^{-\beta E}/Z$ acts as a Boltzmann probability for the initial occupancy-configuration of the two sites. $P(i,n_i;j,n_j) $  is the amplitude for phonon emission or absorption, as the electron system changes from a configuration with $(i,n_i)$, $(j,n_j-1)$ to one with $(i,n_i-1)$, $(j,n_j)$. Up to pre-exponential factors, which we approximate by a uniform value here (set to 1 by a choice of unit of time), $P(i,n_i;j,n_j) $ is given by $N(i,n_i,j,n_j) \equiv \frac{1}{e^{\beta |\Delta E(i,n_i;j,n_j)|} - 1}$ (absorption, $\Delta E>0$)  and $(1 + N(i,n_i,j,n_j))$  for emission, $\Delta E<0 $. The two cases can be combined into a single expression for $N(i,n_i;j,n_j)$:
    
    \begin{equation}
    \tag{A2}
    \label{Planck}
    P(i,n_i;j,n_j)  = \left| \frac{1}{e^{\beta \Delta E(i,n_i;j,n_j)} - 1} \right|
     \end{equation}
  
  Finally, $Z$ is the ``two-site partition function" expressed as 
   \begin{equation*}
    Z = \sum_{n_i,n_j}e^{ -\beta E(n_i,n_j)}
   \end{equation*}  
   where the sum over $n_i$ and $n_j$ runs from 0 to 2.

   A pair-hop from site $i$ to site $j$ corresponds to a resistor between nodes $(i,n_i=2)$ and $(j,n_j=2)$ defined by analogy to Eqn. \ref{transition} as 
   \begin{equation}
   \tag{A3}
   \label{pairtransition}
    \Gamma^0(i,n_i;j,n_j) =  e^{-\frac{2r_{ij}}{\xi}}   \frac{e^{-\beta E(n_i,n_j-2)} }{Z}   P(i,n_i;j,n_j) 
   \end{equation}
 where the variables are the same as defined in Eqn. \ref{transition}, except that $E(n_i,n_j-2)$ is the total energy of the system when sites $i$ and $j$ have occupancies $n_i=2$ and $n_j-2=0$, respectively. The nodes connected by this pair hop, namely $(i,n_i=2)$ and $(j,n_j=2)$, are also connected by a single hop (the last hop described in Table \ref{Resistors}) and we choose the smaller of the two resistance values as the effective resistance between the vertices. 
 
    Note the difference of Eqn. \ref{transition} from a similar expression for the time-averaged rate of transfer given in Ref. \onlinecite{ES1984}. In the latter, for a single hop from $i$ to $j$, the probability of occupancy $n_i=1,n_j=0$ is implemented through a product of the individual occupancy probabilities as $f_i(1-f_j)$, where $f_i$ is the Fermi distribution function, instead of the Boltzmann probability-term $\mathrm{e}^{-\beta E}/Z$ included here. In such an approach, detailed balance in the absence of an electric field can be obtained only by dropping the ``polaron term"  $\frac{e^2}{r_{ij}}$ (particle-hole interaction) when calculating $\Delta E(i,n_i;j,n_j)$ . This simplification leads to a somewhat different value of the Efros-Shklovskii temperature $T_0$ (cf. Eqn. \ref{Efros-ShklovskiiTemperature}) as compared to the treatment used here~\footnote{Private communication with J. Bergli}. However, the particle-hole interaction may play a more vital role in our model, where we include electron pairs: the increased importance arises from the fact  that the polaron term is equal to $\frac{e^2}{r_{ij}}$ for a single hop but 4 times that for a pair hop and thus may be rather significant in the mixed regime favoring pair transport as a whole over pair breaking. 
    
    As a check of our prescription, it can be easily verified that upon barring the polaron term, in the limit of large $U$ where pair formation/transport is hindered, the expression for the transition rate given in Eqn. \ref{transition} completely agrees with the resistor network construction used in Ref. \onlinecite{ES1984} (see also Ref. \onlinecite{Amir2009}).

   In the presence of a weak electric field, one can associate a resistor between the nodes $(i,n_i)$ and $(j,n_j)$ with resistance value given by
   \begin{equation}
   \tag{A4}
   \label{resistance}
    R(i,n_i;j,n_j) = R(j,n_j;i,n_i)= \frac{kT}{e^2 \Gamma^0(i,n_i;j,n_j)},
   \end{equation}
 which is guaranteed to be nondirectional due to detailed balance.

\begin{table}
    \begin{tabular}{|p{2cm}|p{2cm}|c|} 
    \hline
    \multicolumn{2}{|c|}{Initial occupancies} & \multirow{2}{*}{Resistance $R(i,n_i;j,n_j)$}\\
    \cline{1-2}
     Site $i$ & Site $j$ & \\
    \hline
    1 & 0 & $R(i,1;j,1)$\\
    \hline
    1 & 1 & $R(i,1;j,2)$ \\
    \hline
    2 & 0 & $R(i,2;j,1)$ \\
    \hline
    2 & 1 & $R(i,2;j,2)$\\
    \hline     
 \end{tabular}
  \caption{Description of possible single-particle hopping processes between two sites $i$ and $j$ through resistances constructed from nodes of the form $(i,n_i)$ : it is important to note that $n_i$ is the occupancy of site $i$ \emph{before} the hop, while $n_j$ is the occupancy of site $j$ \emph{after} the hop}
  \label{Resistors}
\end{table}

\subsection*{2. Extension to include second order processes}

  The resistor network mapping stated above allows a generalization to include second order processes. Below, we shall describe an extension to include generic two-particle hops and then make approximations to a specific set of two particle hops: namely, ones that involve the formation or distintegration of a pair.
  
\begin{figure}[h]
\centering
  \includegraphics[angle=0, width = 0.45\textwidth]{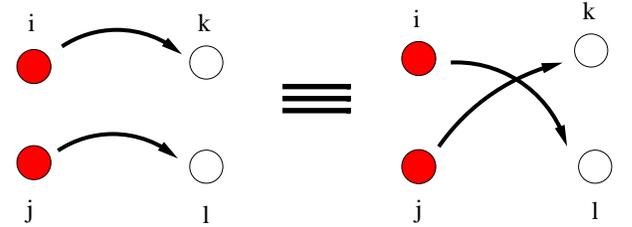}
  \vspace*{10pt}
  \caption{One activated resistance, $R(i,n_i,j,n_j;k,n_k,l,n_l)$, corresponds to the above two hops, as the particles are indistinguishable from each other }
  \label{TwoParticleHop}
\end{figure}
  
   To describe a second-order process, we extend the notion of a node: now, each node is defined by a \emph{pair} of sites with corresponding occupancies as $(i,n_i,j,n_j)$, subject to the constraint $n_i+ n_j\geq 2$.  For a two-particle move from node $(i,n_i,j,n_j)$ to $(k,n_k,l,n_l)$ (note: $n_i$ and $n_j$ are the occupancies of sites $i$ and $j$ \emph{before} the two-particle hop has taken place, while  $n_k$ and $n_l$ are the occupancies of sites $k$ and $l$ \emph{after}), the time-averaged rate of transfer of electrons can be given by the equation 
 \begin{widetext}
  \begin{equation}
   \tag{A5}
   \Gamma^0(i,n_i,j,n_j;k,n_k,l,n_l) = e^{-\frac{2r_{ij;kl}}{\xi} } \frac{e^{ -\beta E(n_i,n_j,n_k-1,n_l-1)}}{Z} P(i,n_i,j,n_j;k,n_k,l,n_l)
  \end{equation}
 \end{widetext}
   Here, $r_{ij;kl}$ is taken as $\min \left[ r_{ik} + r_{jl} , r_{il} + r_{jk}\right]$, keeping in mind the indistinguishability of the hops $(i,n_i,j,n_j) \rightarrow (k,n_k,l,n_l) $ and $(i,n_i,j,n_j) \rightarrow (l,n_l,k,n_k) $ (see figure~\ref{TwoParticleHop}).
   
    Although such a two-particle move potentially includes formation or breaking of pairs, we choose for $\xi$ the localization length of the single-electron wave function in all cases except for pair moves, $(i,n_i,i,n_i)\rightarrow(k,n_k,k,n_k)$ (see Ref. \onlinecite{Bhatt2010} for a discussion of such issues in the context of the Hubbard model). Other than pair hops, the remaining first order processes are treated as a subnetwork of this network of generalized nodes in the following sense --- a single particle-hop from site $(i,n_i)$ to $(j,n_j)$ involves a ``trace" over resistances of the form  $R(i,n_i,k,n_k;j,n_j,k,n_k)$ with the indices $k$ and $n_k$ running over all sites and occupancies respectively. In practice, in the percolation approach, this ``trace" is performed by ``activating" resistances $R(i,n_i,k,n_k;j,n_j,k,n_k)$ for all $k$ and $n_k$ simultaneously for the single-particle hop $(i,n_i) \rightarrow (j,n_j)$ since these resistances all have the same magnitude and are distinguished only by their location in node-space.
    
     The probability for absorbing (or emitting) a phonon with the required energy is in analogy to Eq.~(\ref{Planck}):
   \begin{equation}
    P(i,n_i,j,n_j;k,n_k,l,n_l) = \left| \frac{1}{e^{\beta \Delta E(i,n_i,j,n_j;k,n_k,l,n_l )} - 1} \right|;,\nonumber
   \end{equation}
  and the resistance associated with the link between the nodes is :
    \begin{equation}
    \tag{A6}
    \label{twoparticleresistance}
     R(i,n_i,j,n_j;k,n_k,l,n_l) = \frac{kT}{e^2  \Gamma^0(i,n_i,j,n_j;k,n_k,l,n_l)}
    \end{equation}

 \subsection*{3. Simplified algorithms to include pair breaking/formation}
 
 The  percolation algorithm  with generalized nodes $(i,n_i,j,n_j)$ takes into account all possible second order processes whose number grows roughly as the fourth power of the system size. This constrains the approach to small system sizes (upto a 50 X 50 lattice). Nevertheless, performing the full percolation analysis with the generalized node-network for such small sizes allows comparison with certain approximate networks that we describe below. Once the results are seen to agree well, these latter approximate networks can then be used on larger systems to calculate the effective network-resistance through the percolation approach \cite{Ambegaokar1971}.
 
 The first simplification comes from the sparsity of the generalized resistor network in the sense that low-resistance second order processes (within percolation threshold) involve only small-range hops. As a result, for instance, for a 50 X 50 lattice, while calculating the resistances for second order processes, we can restrict the range of hopping to within 5 sites only. 
 
  Since, for reasons discussed above, we intend to focus on pair formation/disintegration only, another approximation is to retain only those resistances that correspond to this class of second-order processes. In the node-language, these resistances are of the form $R(i,n_i,j,n_j;k,n_k,k,n_k)$ (pair-formation) and $R(k,n_k,k,n_k;i,n_i,j,n_j)$ (pair-breaking). This imposes storage requirements which go as a third power of the system size. 
  
  Moreover, with this simplification, we can revert to using nodes of the form $(i,n_i)$ for first order processes. In effect, we are thus using a mixed definition of nodes, depending on the order of the process, which is unsuitable for a textbook percolation analysis. Let us therefore describe the percolation approach in slightly more technical detail. We activate the resistances in increasing order of magnitude, irrespective of the order of the process (and thus the definition of the nodes connected).
  
   During the percolation analysis (``activating" resistances in ascending order of magnitude till a percolating cluster is obtained), we use the stored second-order resistances to look for an effective `short cut' pair transport in the following way. While activating, say a resistance corresponding to pair breaking originating from site $i$, we check if the lone electrons from this pair breaking are connected through a path of already activated first order single-particle processes to another already activated second-order pair-formation at some other site $j$. If such a path exists, we refer to it as a `short cut' pair-transport. In this case, we treat the nodes $(i,n_i)$ and $(j,n_j)$ as if connected through a pair-hop with the equivalent resistance equal to the last-resistance activated in this `short-cut' path. The advantage of this approximate approach is that the percolation criterion is being applied effectively only to the simpler node-network with nodes of the form $(i,n_i)$ and thus involves a drastic reduction in computation time.
  
   With these two simplifications, it is feasible to study transport for system sizes as high as 200 X 200. This approximate algorithm has been used to check the robustness of the results obtained in the paper to inclusion of the higher-order processes. We find that as anticipated, the resistance for intermediate $U$ (between single and pair-dominated regimes) does decrease due to the extra channels thus included. However, the results described in the main text --- notably the nonmonotonicity of the resistance--- remain present.

\bibliography{TwoCompVRH}

\begin{thebibliography}{54}%
\makeatletter
\providecommand \@ifxundefined [1]{%
 \@ifx{#1\undefined}
}%
\providecommand \@ifnum [1]{%
 \ifnum #1\expandafter \@firstoftwo
 \else \expandafter \@secondoftwo
 \fi
}%
\providecommand \@ifx [1]{%
 \ifx #1\expandafter \@firstoftwo
 \else \expandafter \@secondoftwo
 \fi
}%
\providecommand \natexlab [1]{#1}%
\providecommand \enquote  [1]{``#1''}%
\providecommand \bibnamefont  [1]{#1}%
\providecommand \bibfnamefont [1]{#1}%
\providecommand \citenamefont [1]{#1}%
\providecommand \href@noop [0]{\@secondoftwo}%
\providecommand \href [0]{\begingroup \@sanitize@url \@href}%
\providecommand \@href[1]{\@@startlink{#1}\@@href}%
\providecommand \@@href[1]{\endgroup#1\@@endlink}%
\providecommand \@sanitize@url [0]{\catcode `\\12\catcode `\$12\catcode
  `\&12\catcode `\#12\catcode `\^12\catcode `\_12\catcode `\%12\relax}%
\providecommand \@@startlink[1]{}%
\providecommand \@@endlink[0]{}%
\providecommand \url  [0]{\begingroup\@sanitize@url \@url }%
\providecommand \@url [1]{\endgroup\@href {#1}{\urlprefix }}%
\providecommand \urlprefix  [0]{URL }%
\providecommand \Eprint [0]{\href }%
\providecommand \doibase [0]{http://dx.doi.org/}%
\providecommand \selectlanguage [0]{\@gobble}%
\providecommand \bibinfo  [0]{\@secondoftwo}%
\providecommand \bibfield  [0]{\@secondoftwo}%
\providecommand \translation [1]{[#1]}%
\providecommand \BibitemOpen [0]{}%
\providecommand \bibitemStop [0]{}%
\providecommand \bibitemNoStop [0]{.\EOS\space}%
\providecommand \EOS [0]{\spacefactor3000\relax}%
\providecommand \BibitemShut  [1]{\csname bibitem#1\endcsname}%
\let\auto@bib@innerbib\@empty
\bibitem [{\citenamefont {Hebard}\ and\ \citenamefont
  {Paalanen}(1990)}]{Hebard1990}%
  \BibitemOpen
  \bibfield  {author} {\bibinfo {author} {\bibfnamefont {A.~F.}\ \bibnamefont
  {Hebard}}\ and\ \bibinfo {author} {\bibfnamefont {M.~A.}\ \bibnamefont
  {Paalanen}},\ }\href@noop {} {\bibfield  {journal} {\bibinfo  {journal}
  {Phys. Rev. Lett.}\ }\textbf {\bibinfo {volume} {65}},\ \bibinfo {pages}
  {927} (\bibinfo {year} {1990})}\BibitemShut {NoStop}%
\bibitem [{\citenamefont {Paalanen}\ \emph {et~al.}(1992)\citenamefont
  {Paalanen}, \citenamefont {Hebard},\ and\ \citenamefont
  {Ruel}}]{Paalanen1992}%
  \BibitemOpen
  \bibfield  {author} {\bibinfo {author} {\bibfnamefont {M.~A.}\ \bibnamefont
  {Paalanen}}, \bibinfo {author} {\bibfnamefont {A.~F.}\ \bibnamefont
  {Hebard}}, \ and\ \bibinfo {author} {\bibfnamefont {R.~R.}\ \bibnamefont
  {Ruel}},\ }\href@noop {} {\bibfield  {journal} {\bibinfo  {journal} {Phys.
  Rev. Lett.}\ }\textbf {\bibinfo {volume} {69}},\ \bibinfo {pages} {1604}
  (\bibinfo {year} {1992})}\BibitemShut {NoStop}%
\bibitem [{\citenamefont {Shahar}\ and\ \citenamefont
  {Ovadyahu}(1992)}]{Shahar1992}%
  \BibitemOpen
  \bibfield  {author} {\bibinfo {author} {\bibfnamefont {D.}~\bibnamefont
  {Shahar}}\ and\ \bibinfo {author} {\bibfnamefont {Z.}~\bibnamefont
  {Ovadyahu}},\ }\href@noop {} {\bibfield  {journal} {\bibinfo  {journal}
  {Phys. Rev. B}\ }\textbf {\bibinfo {volume} {46}},\ \bibinfo {pages} {10917}
  (\bibinfo {year} {1992})}\BibitemShut {NoStop}%
\bibitem [{\citenamefont {Gantmakher}\ \emph {et~al.}(1998)\citenamefont
  {Gantmakher}, \citenamefont {Golubkov}, \citenamefont {Dolgopolov},
  \citenamefont {Tsydynzhapov},\ and\ \citenamefont
  {Shashkin}}]{GantmakherInOxpeak}%
  \BibitemOpen
  \bibfield  {author} {\bibinfo {author} {\bibfnamefont {V.~F.}\ \bibnamefont
  {Gantmakher}}, \bibinfo {author} {\bibfnamefont {M.~V.}\ \bibnamefont
  {Golubkov}}, \bibinfo {author} {\bibfnamefont {V.~T.}\ \bibnamefont
  {Dolgopolov}}, \bibinfo {author} {\bibfnamefont {G.~E.}\ \bibnamefont
  {Tsydynzhapov}}, \ and\ \bibinfo {author} {\bibfnamefont {A.~A.}\
  \bibnamefont {Shashkin}},\ }\href@noop {} {\bibfield  {journal} {\bibinfo
  {journal} {JETP Lett.}\ }\textbf {\bibinfo {volume} {68}},\ \bibinfo {pages}
  {337} (\bibinfo {year} {1998})}\BibitemShut {NoStop}%
\bibitem [{\citenamefont {Sambandhamurthy}\ \emph {et~al.}(2004)\citenamefont
  {Sambandhamurthy}, \citenamefont {Engel}, \citenamefont {Johansson},\ and\
  \citenamefont {Shahar}}]{Sambandhamurthy2004}%
  \BibitemOpen
  \bibfield  {author} {\bibinfo {author} {\bibfnamefont {G.}~\bibnamefont
  {Sambandhamurthy}}, \bibinfo {author} {\bibfnamefont {L.}~\bibnamefont
  {Engel}}, \bibinfo {author} {\bibfnamefont {A.}~\bibnamefont {Johansson}}, \
  and\ \bibinfo {author} {\bibfnamefont {D.}~\bibnamefont {Shahar}},\
  }\href@noop {} {\bibfield  {journal} {\bibinfo  {journal} {Phys. Rev. Lett.}\
  }\textbf {\bibinfo {volume} {92}},\ \bibinfo {pages} {107005} (\bibinfo
  {year} {2004})}\BibitemShut {NoStop}%
\bibitem [{\citenamefont {Sambandhamurthy}\ \emph {et~al.}(2005)\citenamefont
  {Sambandhamurthy}, \citenamefont {Engel}, \citenamefont {Johansson},
  \citenamefont {Peled},\ and\ \citenamefont {Shahar}}]{Shahar2005}%
  \BibitemOpen
  \bibfield  {author} {\bibinfo {author} {\bibfnamefont {G.}~\bibnamefont
  {Sambandhamurthy}}, \bibinfo {author} {\bibfnamefont {L.~W.}\ \bibnamefont
  {Engel}}, \bibinfo {author} {\bibfnamefont {A.}~\bibnamefont {Johansson}},
  \bibinfo {author} {\bibfnamefont {E.}~\bibnamefont {Peled}}, \ and\ \bibinfo
  {author} {\bibfnamefont {D.}~\bibnamefont {Shahar}},\ }\href@noop {}
  {\bibfield  {journal} {\bibinfo  {journal} {Phys. Rev. Lett.}\ }\textbf
  {\bibinfo {volume} {94}},\ \bibinfo {pages} {017003} (\bibinfo {year}
  {2005})}\BibitemShut {NoStop}%
\bibitem [{\citenamefont {Nguyen}\ \emph {et~al.}(2009)\citenamefont {Nguyen},
  \citenamefont {Hollen}, \citenamefont {Stewart~Jr.}, \citenamefont
  {Shainline}, \citenamefont {Aijun}, \citenamefont {Xu},\ and\ \citenamefont
  {Valles~Jr.}}]{Valles}%
  \BibitemOpen
  \bibfield  {author} {\bibinfo {author} {\bibfnamefont {H.~Q.}\ \bibnamefont
  {Nguyen}}, \bibinfo {author} {\bibfnamefont {S.~M.}\ \bibnamefont {Hollen}},
  \bibinfo {author} {\bibfnamefont {M.~D.}\ \bibnamefont {Stewart~Jr.}},
  \bibinfo {author} {\bibfnamefont {J.}~\bibnamefont {Shainline}}, \bibinfo
  {author} {\bibfnamefont {Y.}~\bibnamefont {Aijun}}, \bibinfo {author}
  {\bibfnamefont {J.~M.}\ \bibnamefont {Xu}}, \ and\ \bibinfo {author}
  {\bibfnamefont {J.~M.}\ \bibnamefont {Valles~Jr.}},\ }\href@noop {}
  {\bibfield  {journal} {\bibinfo  {journal} {Phys. Rev. Lett.}\ }\textbf
  {\bibinfo {volume} {103}},\ \bibinfo {pages} {157001} (\bibinfo {year}
  {2009})}\BibitemShut {NoStop}%
\bibitem [{\citenamefont {Baturina}\ \emph {et~al.}(2008)\citenamefont
  {Baturina}, \citenamefont {Mironov}, \citenamefont {Vinokur}, \citenamefont
  {Baklanov},\ and\ \citenamefont {Strunk}}]{Baturina2008}%
  \BibitemOpen
  \bibfield  {author} {\bibinfo {author} {\bibfnamefont {T.~I.}\ \bibnamefont
  {Baturina}}, \bibinfo {author} {\bibfnamefont {A.~Y.}\ \bibnamefont
  {Mironov}}, \bibinfo {author} {\bibfnamefont {V.~M.}\ \bibnamefont
  {Vinokur}}, \bibinfo {author} {\bibfnamefont {M.~R.}\ \bibnamefont
  {Baklanov}}, \ and\ \bibinfo {author} {\bibfnamefont {C.}~\bibnamefont
  {Strunk}},\ }\href@noop {} {\bibfield  {journal} {\bibinfo  {journal} {Pis'
  ma v ZhETF}\ }\textbf {\bibinfo {volume} {88}},\ \bibinfo {pages} {867}
  (\bibinfo {year} {2008})}\BibitemShut {NoStop}%
\bibitem [{\citenamefont {Sac\'{e}p\'{e}}(2011)}]{Sacepe2011}%
  \BibitemOpen
  \bibfield  {author} {\bibinfo {author} {\bibfnamefont {B.}~\bibnamefont
  {Sac\'{e}p\'{e}}},\ }\href@noop {} {\bibfield  {journal} {\bibinfo  {journal}
  {Nature Phys.}\ }\textbf {\bibinfo {volume} {7}},\ \bibinfo {pages} {239}
  (\bibinfo {year} {2011})}\BibitemShut {NoStop}%
\bibitem [{\citenamefont {Gantmakher}\ and\ \citenamefont
  {Dolgopolov}(2010)}]{GantmakherDolgopolov}%
  \BibitemOpen
  \bibfield  {author} {\bibinfo {author} {\bibfnamefont {V.~F.}\ \bibnamefont
  {Gantmakher}}\ and\ \bibinfo {author} {\bibfnamefont {V.~T.}\ \bibnamefont
  {Dolgopolov}},\ }\href@noop {} {\bibfield  {journal} {\bibinfo  {journal}
  {Physics-Uspekhi}\ }\textbf {\bibinfo {volume} {53(1)}},\ \bibinfo {pages}
  {1} (\bibinfo {year} {2010})}\BibitemShut {NoStop}%
\bibitem [{\citenamefont {Fisher}(1990)}]{Fisher}%
  \BibitemOpen
  \bibfield  {author} {\bibinfo {author} {\bibfnamefont {M.}~\bibnamefont
  {Fisher}},\ }\href@noop {} {\bibfield  {journal} {\bibinfo  {journal} {Phys.
  Rev. Lett.}\ }\textbf {\bibinfo {volume} {65}},\ \bibinfo {pages} {923}
  (\bibinfo {year} {1990})}\BibitemShut {NoStop}%
\bibitem [{\citenamefont {Finkelstein}(1987)}]{Finkelstein}%
  \BibitemOpen
  \bibfield  {author} {\bibinfo {author} {\bibfnamefont {A.}~\bibnamefont
  {Finkelstein}},\ }\href@noop {} {\bibfield  {journal} {\bibinfo  {journal}
  {JETP Lett.}\ }\textbf {\bibinfo {volume} {45}},\ \bibinfo {pages} {46}
  (\bibinfo {year} {1987})}\BibitemShut {NoStop}%
\bibitem [{\citenamefont {Feigel'man}\ \emph
  {et~al.}(2010{\natexlab{a}})\citenamefont {Feigel'man}, \citenamefont
  {Ioffe}, \citenamefont {Kravtsov},\ and\ \citenamefont
  {Cuevas}}]{Feigel'manKravtsov2010}%
  \BibitemOpen
  \bibfield  {author} {\bibinfo {author} {\bibfnamefont {M.~V.}\ \bibnamefont
  {Feigel'man}}, \bibinfo {author} {\bibfnamefont {L.~B.}\ \bibnamefont
  {Ioffe}}, \bibinfo {author} {\bibfnamefont {V.~E.}\ \bibnamefont {Kravtsov}},
  \ and\ \bibinfo {author} {\bibfnamefont {E.}~\bibnamefont {Cuevas}},\
  }\href@noop {} {\bibfield  {journal} {\bibinfo  {journal} {Annals of
  Physics}\ }\textbf {\bibinfo {volume} {325(7)}},\ \bibinfo {pages} {1390}
  (\bibinfo {year} {2010}{\natexlab{a}})}\BibitemShut {NoStop}%
\bibitem [{\citenamefont {Ghoshal}\ \emph {et~al.}(2001)\citenamefont
  {Ghoshal}, \citenamefont {Randeria},\ and\ \citenamefont
  {Trivedi}}]{Trivedi}%
  \BibitemOpen
  \bibfield  {author} {\bibinfo {author} {\bibfnamefont {A.}~\bibnamefont
  {Ghoshal}}, \bibinfo {author} {\bibfnamefont {M.}~\bibnamefont {Randeria}}, \
  and\ \bibinfo {author} {\bibfnamefont {N.}~\bibnamefont {Trivedi}},\
  }\href@noop {} {\bibfield  {journal} {\bibinfo  {journal} {Phys. Rev. B}\
  }\textbf {\bibinfo {volume} {65}},\ \bibinfo {pages} {014501} (\bibinfo
  {year} {2001})}\BibitemShut {NoStop}%
\bibitem [{\citenamefont {Galitski}\ and\ \citenamefont
  {Larkin}(2001)}]{GalitskiLarkin}%
  \BibitemOpen
  \bibfield  {author} {\bibinfo {author} {\bibfnamefont {V.~M.}\ \bibnamefont
  {Galitski}}\ and\ \bibinfo {author} {\bibfnamefont {A.~I.}\ \bibnamefont
  {Larkin}},\ }\href@noop {} {\bibfield  {journal} {\bibinfo  {journal} {Phys.
  Rev. B}\ }\textbf {\bibinfo {volume} {63}},\ \bibinfo {pages} {174506}
  (\bibinfo {year} {2001})}\BibitemShut {NoStop}%
\bibitem [{\citenamefont {Dubi}\ \emph {et~al.}(2007)\citenamefont {Dubi},
  \citenamefont {Meir},\ and\ \citenamefont {Avishai}}]{Meir}%
  \BibitemOpen
  \bibfield  {author} {\bibinfo {author} {\bibfnamefont {Y.}~\bibnamefont
  {Dubi}}, \bibinfo {author} {\bibfnamefont {Y.}~\bibnamefont {Meir}}, \ and\
  \bibinfo {author} {\bibfnamefont {Y.}~\bibnamefont {Avishai}},\ }\href@noop
  {} {\bibfield  {journal} {\bibinfo  {journal} {Nature}\ }\textbf {\bibinfo
  {volume} {449}},\ \bibinfo {pages} {876} (\bibinfo {year}
  {2007})}\BibitemShut {NoStop}%
\bibitem [{\citenamefont {M\"{u}ller}\ and\ \citenamefont
  {Shklovskii}(2009)}]{MullerShklovskii}%
  \BibitemOpen
  \bibfield  {author} {\bibinfo {author} {\bibfnamefont {M.}~\bibnamefont
  {M\"{u}ller}}\ and\ \bibinfo {author} {\bibfnamefont {B.~I.}\ \bibnamefont
  {Shklovskii}},\ }\href@noop {} {\bibfield  {journal} {\bibinfo  {journal}
  {Phys. Rev. B}\ }\textbf {\bibinfo {volume} {79}},\ \bibinfo {pages} {134504}
  (\bibinfo {year} {2009})}\BibitemShut {NoStop}%
\bibitem [{\citenamefont {Pokrovsky}\ \emph {et~al.}(2010)\citenamefont
  {Pokrovsky}, \citenamefont {Falco},\ and\ \citenamefont
  {Nattermann}}]{Nattermann}%
  \BibitemOpen
  \bibfield  {author} {\bibinfo {author} {\bibfnamefont {V.~L.}\ \bibnamefont
  {Pokrovsky}}, \bibinfo {author} {\bibfnamefont {G.~M.}\ \bibnamefont
  {Falco}}, \ and\ \bibinfo {author} {\bibfnamefont {T.}~\bibnamefont
  {Nattermann}},\ }\href@noop {} {\bibfield  {journal} {\bibinfo  {journal}
  {Phys. Rev. Lett.}\ }\textbf {\bibinfo {volume} {105}},\ \bibinfo {pages}
  {267001} (\bibinfo {year} {2010})}\BibitemShut {NoStop}%
\bibitem [{\citenamefont {Reunchan}\ \emph {et~al.}(2011)\citenamefont
  {Reunchan}, \citenamefont {Zhou}, \citenamefont {Limpijumnong}, \citenamefont
  {Janotti},\ and\ \citenamefont {Van~de Walle}}]{Reunchan2011}%
  \BibitemOpen
  \bibfield  {author} {\bibinfo {author} {\bibfnamefont {P.}~\bibnamefont
  {Reunchan}}, \bibinfo {author} {\bibfnamefont {X.}~\bibnamefont {Zhou}},
  \bibinfo {author} {\bibfnamefont {S.}~\bibnamefont {Limpijumnong}}, \bibinfo
  {author} {\bibfnamefont {A.}~\bibnamefont {Janotti}}, \ and\ \bibinfo
  {author} {\bibfnamefont {C.~G.}\ \bibnamefont {Van~de Walle}},\ }\href@noop
  {} {\bibfield  {journal} {\bibinfo  {journal} {Current Applied Physics}\
  }\textbf {\bibinfo {volume} {11}},\ \bibinfo {pages} {296} (\bibinfo {year}
  {2011})}\BibitemShut {NoStop}%
\bibitem [{\citenamefont {Baker}\ \emph {et~al.}(2010)\citenamefont {Baker},
  \citenamefont {Ormeno}, \citenamefont {Gough}, \citenamefont {Matsushita},\
  and\ \citenamefont {Fisher}}]{FisherStanford}%
  \BibitemOpen
  \bibfield  {author} {\bibinfo {author} {\bibfnamefont {P.~J.}\ \bibnamefont
  {Baker}}, \bibinfo {author} {\bibfnamefont {R.~J.}\ \bibnamefont {Ormeno}},
  \bibinfo {author} {\bibfnamefont {C.~E.}\ \bibnamefont {Gough}}, \bibinfo
  {author} {\bibfnamefont {Y.}~\bibnamefont {Matsushita}}, \ and\ \bibinfo
  {author} {\bibfnamefont {I.~R.}\ \bibnamefont {Fisher}},\ }\href@noop {}
  {\bibfield  {journal} {\bibinfo  {journal} {Phys. Rev. B}\ }\textbf {\bibinfo
  {volume} {81}},\ \bibinfo {pages} {064506} (\bibinfo {year}
  {2010})}\BibitemShut {NoStop}%
\bibitem [{\citenamefont {Schmalian}\ and\ \citenamefont
  {Dzero}(2005)}]{Dzero2005}%
  \BibitemOpen
  \bibfield  {author} {\bibinfo {author} {\bibfnamefont {J.}~\bibnamefont
  {Schmalian}}\ and\ \bibinfo {author} {\bibfnamefont {M.}~\bibnamefont
  {Dzero}},\ }\href@noop {} {\bibfield  {journal} {\bibinfo  {journal} {Phys.
  Rev. Lett.}\ }\textbf {\bibinfo {volume} {94}},\ \bibinfo {pages} {157003}
  (\bibinfo {year} {2005})}\BibitemShut {NoStop}%
\bibitem [{Note1()}]{Note1}%
  \BibitemOpen
  \bibinfo {note} {B. Sac\'{e}p\'{e}, private communication.}\BibitemShut
  {Stop}%
\bibitem [{\citenamefont {Efros}\ and\ \citenamefont
  {Shklovskii}(1975)}]{ES1975}%
  \BibitemOpen
  \bibfield  {author} {\bibinfo {author} {\bibfnamefont {A.~L.}\ \bibnamefont
  {Efros}}\ and\ \bibinfo {author} {\bibfnamefont {B.~I.}\ \bibnamefont
  {Shklovskii}},\ }\href@noop {} {\bibfield  {journal} {\bibinfo  {journal} {J.
  Phys. C}\ }\textbf {\bibinfo {volume} {8}},\ \bibinfo {pages} {49} (\bibinfo
  {year} {1975})}\BibitemShut {NoStop}%
\bibitem [{\citenamefont {Feigel'man}\ \emph
  {et~al.}(2010{\natexlab{b}})\citenamefont {Feigel'man}, \citenamefont
  {Ioffe},\ and\ \citenamefont {M\'{e}zard}}]{Feigel'man2010}%
  \BibitemOpen
  \bibfield  {author} {\bibinfo {author} {\bibfnamefont {M.~V.}\ \bibnamefont
  {Feigel'man}}, \bibinfo {author} {\bibfnamefont {L.~B.}\ \bibnamefont
  {Ioffe}}, \ and\ \bibinfo {author} {\bibfnamefont {M.}~\bibnamefont
  {M\'{e}zard}},\ }\href@noop {} {\bibfield  {journal} {\bibinfo  {journal}
  {Phys. Rev. B}\ }\textbf {\bibinfo {volume} {82}},\ \bibinfo {pages} {184534}
  (\bibinfo {year} {2010}{\natexlab{b}})}\BibitemShut {NoStop}%
\bibitem [{\citenamefont {Dubi}\ \emph {et~al.}(2006)\citenamefont {Dubi},
  \citenamefont {Meir},\ and\ \citenamefont {Avishai}}]{Dubi2006}%
  \BibitemOpen
  \bibfield  {author} {\bibinfo {author} {\bibfnamefont {Y.}~\bibnamefont
  {Dubi}}, \bibinfo {author} {\bibfnamefont {Y.}~\bibnamefont {Meir}}, \ and\
  \bibinfo {author} {\bibfnamefont {Y.}~\bibnamefont {Avishai}},\ }\href@noop
  {} {\bibfield  {journal} {\bibinfo  {journal} {Phys. Rev. B}\ }\textbf
  {\bibinfo {volume} {73}},\ \bibinfo {pages} {054509} (\bibinfo {year}
  {2006})}\BibitemShut {NoStop}%
\bibitem [{\citenamefont {Chen}\ \emph
  {et~al.}(2012{\natexlab{a}})\citenamefont {Chen}, \citenamefont {Skinner},\
  and\ \citenamefont {Shklovskii}}]{Chen2012}%
  \BibitemOpen
  \bibfield  {author} {\bibinfo {author} {\bibfnamefont {T.}~\bibnamefont
  {Chen}}, \bibinfo {author} {\bibfnamefont {B.}~\bibnamefont {Skinner}}, \
  and\ \bibinfo {author} {\bibfnamefont {B.~I.}\ \bibnamefont {Shklovskii}},\
  }\href@noop {} {\bibfield  {journal} {\bibinfo  {journal} {arXiv :
  1204.4935v1}\ } (\bibinfo {year} {2012}{\natexlab{a}})}\BibitemShut {NoStop}%
\bibitem [{\citenamefont {Chen}\ \emph
  {et~al.}(2012{\natexlab{b}})\citenamefont {Chen}, \citenamefont {Skinner},\
  and\ \citenamefont {Shklovskii}}]{Chen2012a}%
  \BibitemOpen
  \bibfield  {author} {\bibinfo {author} {\bibfnamefont {T.}~\bibnamefont
  {Chen}}, \bibinfo {author} {\bibfnamefont {B.}~\bibnamefont {Skinner}}, \
  and\ \bibinfo {author} {\bibfnamefont {B.~I.}\ \bibnamefont {Shklovskii}},\
  }\href@noop {} {\bibfield  {journal} {\bibinfo  {journal}
  {arXiv:1203.3889v2}\ } (\bibinfo {year} {2012}{\natexlab{b}})}\BibitemShut
  {NoStop}%
\bibitem [{\citenamefont {Efros}\ and\ \citenamefont
  {Shklovskii}(1984)}]{ES1984}%
  \BibitemOpen
  \bibfield  {author} {\bibinfo {author} {\bibfnamefont {A.~L.}\ \bibnamefont
  {Efros}}\ and\ \bibinfo {author} {\bibfnamefont {B.~I.}\ \bibnamefont
  {Shklovskii}},\ }\href@noop {} {\emph {\bibinfo {title} {Electronic
  properties of doped semiconductors}}}\ (\bibinfo  {publisher} {Springer
  Berlin},\ \bibinfo {year} {1984})\BibitemShut {NoStop}%
\bibitem [{\citenamefont {Gangopadhyay}\ \emph {et~al.}()\citenamefont
  {Gangopadhyay}, \citenamefont {Galitski},\ and\ \citenamefont
  {M\"{u}ller}}]{Gangopadhyay}%
  \BibitemOpen
  \bibfield  {author} {\bibinfo {author} {\bibfnamefont {A.}~\bibnamefont
  {Gangopadhyay}}, \bibinfo {author} {\bibfnamefont {V.~M.}\ \bibnamefont
  {Galitski}}, \ and\ \bibinfo {author} {\bibfnamefont {M.}~\bibnamefont
  {M\"{u}ller}},\ }\href@noop {} {\bibinfo  {journal} {to be published}\
  }\BibitemShut {NoStop}%
\bibitem [{\citenamefont {Pikus}\ and\ \citenamefont
  {Efros}(1994)}]{EfrosPikus}%
  \BibitemOpen
\bibfield  {journal} {  }\bibfield  {author} {\bibinfo {author} {\bibfnamefont
  {F.~G.}\ \bibnamefont {Pikus}}\ and\ \bibinfo {author} {\bibfnamefont
  {A.~L.}\ \bibnamefont {Efros}},\ }\href@noop {} {\bibfield  {journal}
  {\bibinfo  {journal} {Phys. Rev. Lett.}\ }\textbf {\bibinfo {volume} {73}},\
  \bibinfo {pages} {3014} (\bibinfo {year} {1994})}\BibitemShut {NoStop}%
\bibitem [{\citenamefont {M\"{u}ller}\ and\ \citenamefont
  {Pankov}(2007)}]{MullerPankov07}%
  \BibitemOpen
  \bibfield  {author} {\bibinfo {author} {\bibfnamefont {M.}~\bibnamefont
  {M\"{u}ller}}\ and\ \bibinfo {author} {\bibfnamefont {S.}~\bibnamefont
  {Pankov}},\ }\href@noop {} {\bibfield  {journal} {\bibinfo  {journal} {Phys.
  Rev. B}\ }\textbf {\bibinfo {volume} {75}},\ \bibinfo {pages} {144201}
  (\bibinfo {year} {2007})}\BibitemShut {NoStop}%
\bibitem [{\citenamefont {M\"{o}bius}\ \emph {et~al.}(2009)\citenamefont
  {M\"{o}bius}, \citenamefont {Karmann},\ and\ \citenamefont
  {Schreiber}}]{Mobius2009}%
  \BibitemOpen
  \bibfield  {author} {\bibinfo {author} {\bibfnamefont {A.}~\bibnamefont
  {M\"{o}bius}}, \bibinfo {author} {\bibfnamefont {P.}~\bibnamefont {Karmann}},
  \ and\ \bibinfo {author} {\bibfnamefont {M.}~\bibnamefont {Schreiber}},\
  }\href@noop {} {\bibfield  {journal} {\bibinfo  {journal} {J. Phys.: Conf.
  Ser.}\ }\textbf {\bibinfo {volume} {150}},\ \bibinfo {pages} {022057}
  (\bibinfo {year} {2009})}\BibitemShut {NoStop}%
\bibitem [{\citenamefont {Levin}\ \emph {et~al.}(1987)\citenamefont {Levin},
  \citenamefont {Nguen}, \citenamefont {Shklovskii},\ and\ \citenamefont
  {Efros}}]{ES1987}%
  \BibitemOpen
  \bibfield  {author} {\bibinfo {author} {\bibfnamefont {E.~I.}\ \bibnamefont
  {Levin}}, \bibinfo {author} {\bibfnamefont {V.~L.}\ \bibnamefont {Nguen}},
  \bibinfo {author} {\bibfnamefont {B.~I.}\ \bibnamefont {Shklovskii}}, \ and\
  \bibinfo {author} {\bibfnamefont {A.~L.}\ \bibnamefont {Efros}},\ }\href@noop
  {} {\bibfield  {journal} {\bibinfo  {journal} {Sov. Phys. JETP}\ }\textbf
  {\bibinfo {volume} {65}},\ \bibinfo {pages} {842} (\bibinfo {year}
  {1987})}\BibitemShut {NoStop}%
\bibitem [{\citenamefont {Baranovskii}\ \emph {et~al.}(1979)\citenamefont
  {Baranovskii}, \citenamefont {Efros}, \citenamefont {Gelmont},\ and\
  \citenamefont {Shklovskii}}]{Baranovskii1979}%
  \BibitemOpen
  \bibfield  {author} {\bibinfo {author} {\bibfnamefont {S.~D.}\ \bibnamefont
  {Baranovskii}}, \bibinfo {author} {\bibfnamefont {A.}~\bibnamefont {Efros}},
  \bibinfo {author} {\bibfnamefont {B.~L.}\ \bibnamefont {Gelmont}}, \ and\
  \bibinfo {author} {\bibfnamefont {B.~I.}\ \bibnamefont {Shklovskii}},\
  }\href@noop {} {\bibfield  {journal} {\bibinfo  {journal} {J. Phys. C}\
  }\textbf {\bibinfo {volume} {12}},\ \bibinfo {pages} {1023} (\bibinfo {year}
  {1979})}\BibitemShut {NoStop}%
\bibitem [{\citenamefont {Goethe}\ and\ \citenamefont
  {Palassini}()}]{Palassini11}%
  \BibitemOpen
  \bibfield  {author} {\bibinfo {author} {\bibfnamefont {M.}~\bibnamefont
  {Goethe}}\ and\ \bibinfo {author} {\bibfnamefont {M.}~\bibnamefont
  {Palassini}},\ }\href@noop {} {\bibinfo  {journal} {to be published}\
  }\BibitemShut {NoStop}%
\bibitem [{\citenamefont {Bardalen}\ \emph {et~al.}(2012)\citenamefont
  {Bardalen}, \citenamefont {Bergli},\ and\ \citenamefont
  {Galperin}}]{Galperin2012}%
  \BibitemOpen
\bibfield  {journal} {  }\bibfield  {author} {\bibinfo {author} {\bibfnamefont
  {E.}~\bibnamefont {Bardalen}}, \bibinfo {author} {\bibfnamefont
  {J.}~\bibnamefont {Bergli}}, \ and\ \bibinfo {author} {\bibfnamefont
  {Y.}~\bibnamefont {Galperin}},\ }\href@noop {} {\bibfield  {journal}
  {\bibinfo  {journal} {Phys. Rev. B}\ }\textbf {\bibinfo {volume} {85}},\
  \bibinfo {pages} {155206} (\bibinfo {year} {2012})}\BibitemShut {NoStop}%
\bibitem [{\citenamefont {Kamimura}\ and\ \citenamefont
  {Kurobe}(1982)}]{Ovadyahu-Kamimuraeffect}%
  \BibitemOpen
  \bibfield  {author} {\bibinfo {author} {\bibfnamefont {H.}~\bibnamefont
  {Kamimura}}\ and\ \bibinfo {author} {\bibfnamefont {A.}~\bibnamefont
  {Kurobe}},\ }\href@noop {} {\bibfield  {journal} {\bibinfo  {journal} {J.
  Phys. Soc. Japan}\ }\textbf {\bibinfo {volume} {51}},\ \bibinfo {pages}
  {1904} (\bibinfo {year} {1982})}\BibitemShut {NoStop}%
\bibitem [{\citenamefont {Vaknin}\ \emph {et~al.}(1996)\citenamefont {Vaknin},
  \citenamefont {Frydman}, \citenamefont {Ovadyahu},\ and\ \citenamefont
  {Pollak}}]{Vaknin96}%
  \BibitemOpen
  \bibfield  {author} {\bibinfo {author} {\bibfnamefont {A.}~\bibnamefont
  {Vaknin}}, \bibinfo {author} {\bibfnamefont {A.}~\bibnamefont {Frydman}},
  \bibinfo {author} {\bibfnamefont {Z.}~\bibnamefont {Ovadyahu}}, \ and\
  \bibinfo {author} {\bibfnamefont {M.}~\bibnamefont {Pollak}},\ }\href@noop {}
  {\bibfield  {journal} {\bibinfo  {journal} {Phys. Rev. B}\ }\textbf {\bibinfo
  {volume} {54}},\ \bibinfo {pages} {13604} (\bibinfo {year}
  {1996})}\BibitemShut {NoStop}%
\bibitem [{\citenamefont {Sachdev}(2000)}]{Sachdev2000}%
  \BibitemOpen
  \bibfield  {author} {\bibinfo {author} {\bibfnamefont {S.}~\bibnamefont
  {Sachdev}},\ }\href@noop {} {\emph {\bibinfo {title} {Quantum Phase
  Transitions}}}\ (\bibinfo  {publisher} {Cambridge University Press},\
  \bibinfo {year} {2000})\BibitemShut {NoStop}%
\bibitem [{\citenamefont {Schechter}\ and\ \citenamefont
  {Stamp}()}]{SchechterStamp2009}%
  \BibitemOpen
  \bibfield  {author} {\bibinfo {author} {\bibfnamefont {M.}~\bibnamefont
  {Schechter}}\ and\ \bibinfo {author} {\bibfnamefont {P.}~\bibnamefont
  {Stamp}},\ }\href@noop {} {\ }\bibinfo {note} {Condmat:
  0910.1283}\BibitemShut {NoStop}%
\bibitem [{\citenamefont {Miller}\ and\ \citenamefont
  {Abrahams}(1960)}]{MillerAbrahams1960}%
  \BibitemOpen
  \bibfield  {author} {\bibinfo {author} {\bibfnamefont {A.}~\bibnamefont
  {Miller}}\ and\ \bibinfo {author} {\bibfnamefont {E.}~\bibnamefont
  {Abrahams}},\ }\href@noop {} {\bibfield  {journal} {\bibinfo  {journal}
  {Phys. Rev.}\ }\textbf {\bibinfo {volume} {103}},\ \bibinfo {pages} {745}
  (\bibinfo {year} {1960})}\BibitemShut {NoStop}%
\bibitem [{\citenamefont {Amir}\ \emph {et~al.}(2009)\citenamefont {Amir},
  \citenamefont {Oreg},\ and\ \citenamefont {Imry}}]{Amir2009}%
  \BibitemOpen
  \bibfield  {author} {\bibinfo {author} {\bibfnamefont {A.}~\bibnamefont
  {Amir}}, \bibinfo {author} {\bibfnamefont {Y.}~\bibnamefont {Oreg}}, \ and\
  \bibinfo {author} {\bibfnamefont {Y.}~\bibnamefont {Imry}},\ }\href@noop {}
  {\bibfield  {journal} {\bibinfo  {journal} {Phys. Rev. B}\ }\textbf {\bibinfo
  {volume} {80}},\ \bibinfo {pages} {245214} (\bibinfo {year}
  {2009})}\BibitemShut {NoStop}%
\bibitem [{\citenamefont {Ambegaokar}\ \emph {et~al.}(1971)\citenamefont
  {Ambegaokar}, \citenamefont {Halperin},\ and\ \citenamefont
  {Langer}}]{Ambegaokar1971}%
  \BibitemOpen
  \bibfield  {author} {\bibinfo {author} {\bibfnamefont {V.}~\bibnamefont
  {Ambegaokar}}, \bibinfo {author} {\bibfnamefont {B.}~\bibnamefont
  {Halperin}}, \ and\ \bibinfo {author} {\bibfnamefont {J.}~\bibnamefont
  {Langer}},\ }\href@noop {} {\bibfield  {journal} {\bibinfo  {journal} {Phys.
  Rev. B}\ }\textbf {\bibinfo {volume} {4}},\ \bibinfo {pages} {2612} (\bibinfo
  {year} {1971})}\BibitemShut {NoStop}%
\bibitem [{\citenamefont {Zhao}\ \emph {et~al.}(1991)\citenamefont {Zhao},
  \citenamefont {Spivak}, \citenamefont {Gelfand},\ and\ \citenamefont
  {Feng}}]{Zhao1991}%
  \BibitemOpen
  \bibfield  {author} {\bibinfo {author} {\bibfnamefont {H.}~\bibnamefont
  {Zhao}}, \bibinfo {author} {\bibfnamefont {B.}~\bibnamefont {Spivak}},
  \bibinfo {author} {\bibfnamefont {M.}~\bibnamefont {Gelfand}}, \ and\
  \bibinfo {author} {\bibfnamefont {S.}~\bibnamefont {Feng}},\ }\href@noop {}
  {\bibfield  {journal} {\bibinfo  {journal} {Phys. Rev. B}\ }\textbf {\bibinfo
  {volume} {44}},\ \bibinfo {pages} {10760} (\bibinfo {year}
  {1991})}\BibitemShut {NoStop}%
\bibitem [{\citenamefont {M\"{u}ller}(2011)}]{Mueller}%
  \BibitemOpen
  \bibfield  {author} {\bibinfo {author} {\bibfnamefont {M.}~\bibnamefont
  {M\"{u}ller}},\ }\href@noop {} {\bibfield  {journal} {\bibinfo  {journal}
  {arXiv : 1109.0245v1}\ } (\bibinfo {year} {2011})}\BibitemShut {NoStop}%
\bibitem [{\citenamefont {Anderson}(1959)}]{Anderson59}%
  \BibitemOpen
  \bibfield  {author} {\bibinfo {author} {\bibfnamefont {P.~W.}\ \bibnamefont
  {Anderson}},\ }\href@noop {} {\bibfield  {journal} {\bibinfo  {journal}
  {J.Phys. Chem. Solids}\ }\textbf {\bibinfo {volume} {11}},\ \bibinfo {pages}
  {26} (\bibinfo {year} {1959})}\BibitemShut {NoStop}%
\bibitem [{\citenamefont {Ma}\ and\ \citenamefont {Lee}(1985)}]{MaLee85}%
  \BibitemOpen
  \bibfield  {author} {\bibinfo {author} {\bibfnamefont {M.}~\bibnamefont
  {Ma}}\ and\ \bibinfo {author} {\bibfnamefont {P.~A.}\ \bibnamefont {Lee}},\
  }\href@noop {} {\bibfield  {journal} {\bibinfo  {journal} {Phys. Rev. B}\
  }\textbf {\bibinfo {volume} {32}},\ \bibinfo {pages} {5658} (\bibinfo {year}
  {1985})}\BibitemShut {NoStop}%
\bibitem [{\citenamefont {Armitage}\ \emph {et~al.}(2004)\citenamefont
  {Armitage}, \citenamefont {Helgren},\ and\ \citenamefont
  {Gruner}}]{Armitage}%
  \BibitemOpen
  \bibfield  {author} {\bibinfo {author} {\bibfnamefont {N.~P.}\ \bibnamefont
  {Armitage}}, \bibinfo {author} {\bibfnamefont {E.}~\bibnamefont {Helgren}}, \
  and\ \bibinfo {author} {\bibfnamefont {G.}~\bibnamefont {Gruner}},\
  }\href@noop {} {\bibfield  {journal} {\bibinfo  {journal} {Phys. Rev. B.}\
  }\textbf {\bibinfo {volume} {69}},\ \bibinfo {pages} {014201} (\bibinfo
  {year} {2004})}\BibitemShut {NoStop}%
\bibitem [{\citenamefont {Massey}\ and\ \citenamefont {Lee}(1995)}]{MasseyLee}%
  \BibitemOpen
  \bibfield  {author} {\bibinfo {author} {\bibfnamefont {J.~G.}\ \bibnamefont
  {Massey}}\ and\ \bibinfo {author} {\bibfnamefont {M.}~\bibnamefont {Lee}},\
  }\href@noop {} {\bibfield  {journal} {\bibinfo  {journal} {Phys. Rev. Lett.}\
  }\textbf {\bibinfo {volume} {75}},\ \bibinfo {pages} {4266} (\bibinfo {year}
  {1995})}\BibitemShut {NoStop}%
\bibitem [{\citenamefont {Vaknin}\ \emph {et~al.}(2002)\citenamefont {Vaknin},
  \citenamefont {Ovadyahu},\ and\ \citenamefont {Pollak}}]{electronglass}%
  \BibitemOpen
  \bibfield  {author} {\bibinfo {author} {\bibfnamefont {A.}~\bibnamefont
  {Vaknin}}, \bibinfo {author} {\bibfnamefont {Z.}~\bibnamefont {Ovadyahu}}, \
  and\ \bibinfo {author} {\bibfnamefont {M.}~\bibnamefont {Pollak}},\
  }\href@noop {} {\bibfield  {journal} {\bibinfo  {journal} {Phys. Rev. B}\
  }\textbf {\bibinfo {volume} {65}},\ \bibinfo {pages} {134208} (\bibinfo
  {year} {2002})}\BibitemShut {NoStop}%
\bibitem [{\citenamefont {Lebanon}\ and\ \citenamefont
  {M\"{u}ller}(2005)}]{Lebanon2005}%
  \BibitemOpen
  \bibfield  {author} {\bibinfo {author} {\bibfnamefont {E.}~\bibnamefont
  {Lebanon}}\ and\ \bibinfo {author} {\bibfnamefont {M.}~\bibnamefont
  {M\"{u}ller}},\ }\href@noop {} {\bibfield  {journal} {\bibinfo  {journal}
  {Phys. Rev. B}\ }\textbf {\bibinfo {volume} {72}},\ \bibinfo {pages} {174202}
  (\bibinfo {year} {2005})}\BibitemShut {NoStop}%
\bibitem [{\citenamefont {Naaman}\ \emph {et~al.}(2001)\citenamefont {Naaman},
  \citenamefont {Tezier},\ and\ \citenamefont {Dynes}}]{Naaman2001}%
  \BibitemOpen
  \bibfield  {author} {\bibinfo {author} {\bibfnamefont {O.}~\bibnamefont
  {Naaman}}, \bibinfo {author} {\bibfnamefont {W.}~\bibnamefont {Tezier}}, \
  and\ \bibinfo {author} {\bibfnamefont {R.~C.}\ \bibnamefont {Dynes}},\
  }\href@noop {} {\bibfield  {journal} {\bibinfo  {journal} {Phys. Rev. Lett.}\
  }\textbf {\bibinfo {volume} {87}},\ \bibinfo {pages} {097004} (\bibinfo
  {year} {2001})}\BibitemShut {NoStop}%
\bibitem [{Note2()}]{Note2}%
  \BibitemOpen
  \bibinfo {note} {Private communication with J. Bergli}\BibitemShut {NoStop}%
\bibitem [{\citenamefont {Bhatt}\ and\ \citenamefont
  {Nielsen}(2010)}]{Bhatt2010}%
  \BibitemOpen
  \bibfield  {author} {\bibinfo {author} {\bibfnamefont {R.~N.}\ \bibnamefont
  {Bhatt}}\ and\ \bibinfo {author} {\bibfnamefont {E.}~\bibnamefont
  {Nielsen}},\ }\href@noop {} {\bibfield  {journal} {\bibinfo  {journal} {Phys.
  Rev. B}\ }\textbf {\bibinfo {volume} {82}},\ \bibinfo {pages} {195117}
  (\bibinfo {year} {2010})}\BibitemShut {NoStop}%
\end{thebibliography}%

\end{document}